\documentclass[12pt]{article}%
\usepackage{setspace}
\usepackage[left=1in,right=1in,top=1.5in,bottom=1.5in]{geometry}
\usepackage{graphicx}%
\usepackage{multirow}%
\usepackage{amsmath,amssymb,amsfonts, authblk}%
\usepackage{amsthm}%
\usepackage{xcolor}%
\usepackage{booktabs}%
\usepackage{bm, color}
\usepackage{xcolor}
\usepackage{natbib}
\usepackage{mathtools}
\usepackage{placeins}

\newtheorem{proposition}{Proposition}
\newtheorem{theorem}{Theorem}
\usepackage{threeparttable}
\usepackage{xspace}
\usepackage{hyperref}
\usepackage{caption} 
\def\LAD{LAD\xspace}
\def\SIR{SIR\xspace}
\def\SAVE{SAVE\xspace}
\def\SDR{SDR\xspace}
\def\T{\textrm{T}}
\def\bX{\mathbf{X}}
\def\bbeta{\bm{\beta}}

\def\bB{\mathbf{B}}
\def\bI{\mathbf{I}}
\def\bZ{\mathbf{Z}}
\def\bW{\mathbf{W}}
\def\bU{\mathbf{U}}
\def\bR{\mathbf{R}}
\def\bV{\mathbf{V}}
\def\bv{\bm{v}}
\def\bA{\mathbf{A}}
\def\bT{\mathbf{T}}
\def\bH{\mathbf{H}}
\def\bK{\mathbf{K}}
\def\bD{\mathbf{D}}

\def\bSigma{\bm{\Sigma}}
\def\bPsi{\bm{\Psi}}
\def\bDelta{\bm{\Delta}}
\def\bPhi{\bm{\Phi}}
\newcommand{\bL}{\mathbf{L}}
\def\bG{\mathbf{G}}
\def\bmu{\bm{\mu}}

\newcommand{\commln}[1]{{\leavevmode\color{black}#1}}
\newcommand{\bfit}[1]{{\textbf{#1}}}
\begin{document}

\title{\vspace{-5em}Likelihood-based surrogate dimension reduction}

\author[1]{Linh H. Nghiem\thanks{Corresponding author: linh.nghiem@sydney.edu.au}} 
\author[2]{Francis K.C. Hui}
\author[1,3]{Samuel M{\"u}ller}
\author[2]{A.H.Welsh}

\affil[1]{School of Mathematics and Statistics, University of Sydney, Sydney, Australia}
\affil[2]{Research School of Finance, Actuarial Studies and Statistics, The Australian National University, Canberra, Australia}
\affil[2]{School of Mathematical and Physical Sciences, Macquarie University, Sydney, Australia}
\date{}
\maketitle
\abstract{We consider the problem of surrogate sufficient dimension reduction, that is, estimating the central subspace of a regression model, when the covariates are contaminated by measurement error. When no measurement error is present, a likelihood-based dimension reduction method that relies on maximizing the likelihood of a Gaussian inverse regression model on the Grassmann manifold is well-known to have superior performance to traditional inverse moment methods. We propose two likelihood-based estimators for the central subspace in measurement error settings, which make different adjustments to the observed surrogates. Both estimators are computed based on maximizing objective functions on the Grassmann manifold and are shown to consistently recover the true central subspace. When the central subspace is assumed to depend on only a few covariates, we further propose to augment the likelihood function with a penalty term that induces sparsity on the Grassmann manifold to obtain sparse estimators. \textcolor{black}{The resulting objective function has a closed-form Riemann gradient which facilitates efficient computation of the penalized estimator. We leverage the state-of-the-art trust region algorithm on the Grassmann manifold to compute the proposed estimators efficiently.} Simulation studies and a data application demonstrate the proposed likelihood-based estimators perform better than inverse moment-based estimators in terms of both estimation and variable selection accuracy.}


\thispagestyle{empty}
\newpage
\section{Introduction}

In a regression setting with an outcome $y \in \mathbb{R}$ and a covariate vector $\bX \in \mathbb{R}^p$, sufficient dimension reduction (SDR) refers to a class of methods that express the outcome as a few linear combinations of $\bX$ \citep{li2018sufficient}. In other words, \SDR aims to estimate the matrix $\bB \in \mathbb{R}^{p \times d}$ such that $y \perp \bX \mid \bB^\T \bX$, where $d \ll p $. As the matrix $\bB$ is generally not unique, the estimation target in \SDR is the central subspace $\mathcal{S}_{y \vert \bX}$, defined as the intersection of all the subspaces spanned by the columns of $\bB$ satisfying the above conditional independence condition. This central subspace is unique under mild conditions \citep{glaws2020inverse}, and is characterized by the projection matrix associated with $\bB$.

This article focuses on the problem of estimating the central subspace $\mathcal{S}_{y \vert \bX}$ when the covariates $\bX$ are measured with error. This problem is known in the literature as surrogate sufficient dimension reduction \citep{li2007surrogate}. Instead of observing a sample for $\bX$, we observe a sample of $\bW$, which are surrogates related to $\bX$ by the classical additive measurement error model $\bW = \bX + \bU$, where $\bU$ is a vector of measurement errors independent of $\bX$ and which follows a $p$-variate Gaussian distribution with mean zero and covariance matrix $\bSigma_u$. \commln{While more general measurement error models have also been proposed in the literature e.g., $\bW = \mathbf{A}\bX + \bU$ for $\mathbf{A} \in \mathbb{R}^{r \times p}$ with $r \geq p$ \citep{carroll1992measurement, li2007surrogate, zhang2014surrogate}, the classical measurement error model with $\mathbf{A} = \mathbf{I}_p$ is still the most widely used in practice \citep[see][for some recent examples]{grace2021handbook, chenyi2022, chen2023noising}, and is the focus of this paper.} Throughout the article we assume $\bSigma_u$ is known; in practice, this covariance matrix may be estimated from auxiliary data, such as replicate observations, before subsequent analyses are conducted \citep{carroll2006measurement}. Our observed data thus consists of $n$ pairs $(y_i, \bW_i^\T)$ with $\bW_i = \bX_i + \bU_i$, $i=1,\ldots, n$.

When the true covariates  $\bX_i$ are observed, there exists a vast literature on how to estimate the central subspace; see  \citet{li2018sufficient} for an overview. These include traditional inverse moment-based methods such as sliced inverse regression \citep{li1991sliced} and sliced average variance estimation \citep{dennis2000save}, forward regression methods such as minimum average variance estimation \citep{xia2002adaptive} and directional regression \citep{li2007directional}, and inverse regression methods such as principal fitted components \citep{cook2008principal} and likelihood-acquired directions \citep[\LAD,][]{cook2009likelihood}. Recent literature has also expanded these methods into more complex settings, such as high-dimensional data \citep[e.g.,][]{lin2017optimality, qian2019sparse} and longitudinal data \citep[e.g.,][]{hui2022sufficient}. Among the estimators discussed above, the \LAD method developed by \citet{cook2009likelihood} has a unique advantage in that it is constructed from a well-defined Gaussian likelihood function and, as such, inherits the optimality properties of likelihood theory. \commln{Also, while the conditional Gaussianity assumption appears unnatural, \citet{cook2009likelihood} show that the \LAD estimator has superior performance even when Gaussianity does not hold; roughly speaking, the conditional normality plays the role of a ``working model.''} In this article, we focus on adapting \LAD to the surrogate \SDR problem.

The problem of surrogate \SDR was first considered in \citet{carroll1992measurement}, who showed that $\mathcal{S}_{y \vert \bX}$ can be estimated by performing ordinary least squares, or sliced inverse regression, of the response on the adjusted surrogate $\bX^* = \bR\bW$, where $\bR = \bSigma_{xw}\bSigma_{w}^{-1} = (\bSigma_{w}-\bSigma_u)\bSigma_{w}^{-1}$. Here, $\bSigma_{xw} = \text{Cov}(\bX,\bW)$ denotes the covariance matrix of $\bX$ and $\bW$ and $\bSigma_{w} = \text{Var}(\bW)$  denotes the variance-covariance matrix of $\bW$. When $\bSigma_u$ is known, these adjusted surrogates can be computed by replacing $\bSigma_w$ with an appropriate estimator from the sample. This underlying idea was expanded into a broader, invariance law by \citet{li2007surrogate}, who prove that if both $\bX$ and $\bU$ follow multivariate Gaussian distributions, then  $\mathcal{S}_{y \vert {\bX^*}} = \mathcal{S}_{y \vert \bX} $. As a result, under the assumption of Gaussianity of both $\bX$ and $\bU$, any consistent \SDR method applied to $y$ and the adjusted surrogate ${\bX}^*$ is also consistent. If $\bX$ is non-Gaussian, this relationship is maintained if $\bB^\T \bX$ is approximately Gaussian. For instance, this is achieved when $p \to \infty$ and each column of $\bB$ is dense i.e., the majority of elements in each column are non-zero. On the other hand, when the number of covariates is large, it is often assumed that the central subspace is sparse \citep[i.e., it depends only on a few covariates][]{Linsparse}, and so this Gaussian approximation may not be realistic or relevant in practice. Elsewhere, \citet{zhang2012dimension} examines the surrogate \SDR problem when both the response and the covariates are distorted by multiplicative measurement errors, while \citet{chenyi2022} proposed a \SDR method for survival data when the response is censored and the covariates are measured with error at the same time. Neither of these addresses the sparsity of the central subspace when $p$ is large,  and more broadly there is little direct research on inverse regression methods for \SDR under measurement error.  

In this paper, we propose two likelihood-based estimators of the central subspace $\mathcal{S}_{y\vert \bX}$ from the surrogate data $(y_i, \bW_i^\T)$. For the first estimator, we directly follow the approach of \LAD and model the inverse predictor $\bX \mid y$ as following a Gaussian distribution, where the existence of the central subspace imposes multiple constraints on the parameters and the covariates are subject to additive measurement errors. Measurement error is then incorporated into this model in a straightforward manner. We construct a likelihood-based function for the semi-orthogonal bases of the central subspace, and show that maximizing this function requires solving an optimization problem on the Grassmann manifold. For the second estimator, we apply the \LAD approach on the adjusted surrogate $\bX^*$.  \commln{Although the two estimators make different adjustments to the adjusted surrogates, we show that these two estimators are asymptotically equivalent in terms of estimating the true central subspace.} \textcolor{black}{Furthermore, we propose a sparse estimator for the central subspace by augmenting the likelihood function with a penalty term that regularizes the elements of the corresponding projection matrix. The resulting objective function has a closed-form Riemann gradient, which facilitates efficient computation of the penalized estimator.} Simulation studies and an application to a National Health and Nutrition Examination Survey from the United States demonstrate that the performance of the proposed likelihood-based estimators is superior to several common \commln{inverse moment-based} estimators \commln{in terms of both estimating the central subspace and variable selection accuracy.} 

The rest of the paper is organized as follows: In Section 2, we briefly review the \LAD estimator of the central subspace, which is built upon a Gaussian inverse regression model. Then we incorporate measurement errors and discuss the properties of the corresponding model. We propose maximum likelihood estimators of the central subspace from the surrogate data in Section 3. Section 4 presents simulation studies to demonstrate the performance of the proposed estimators, while Section 5 illustrates the methodology in a data application. Finally, Section 6 contains some concluding remarks.

\section{A Gaussian inverse regression model with measurement errors}
We first review the \LAD method proposed by \citet{cook2009likelihood}, when the true covariate vector $\bX$ is observed. \LAD models the inverse predictor $\bX \mid y$ as having a multivariate Gaussian distribution with mean $\bmu_y$ and covariance $\bDelta_y$ i.e., $\bX \mid y \sim N(\bmu_y, \bDelta_y)$. Importantly, the existence of the central subspace $\mathcal{S}_{y \vert \bX}$ imposes some constraints on the parameters $\bmu_y$ and $\bDelta_y$. Let $\bmu = \text{E}(\bX)$, $\bDelta = \text{E}(\bDelta_y) = \text{E}\left\{\text{Var}(\bX \mid y) \right\}$, and let $\bPsi \in \mathbb{R}^{p \times d}$ be a semi-orthogonal basis matrix for $\mathcal{S}_{y \vert \bX}$ and $\bPsi_0 \in \mathbb{R}^{p \times (p-d)} $ be a matrix, such that $(\bPsi, \bPsi_0) \in \mathbb{R}^{p \times p}$ is an orthogonal matrix. Then $\mathcal{S}_{y \vert \bX}$ is a central subspace if and only if the two following  conditions are satisfied:
\begin{equation}
\begin{aligned}
& (i)~ \bPsi^\T \bX \mid y \sim N(\bPsi^\T \bmu + \bPsi^\T  \bDelta \bPsi \bv_y, \bPsi^\T \bDelta_y \bPsi), \\ & (ii) ~ \bPsi_0^\T \bX \mid (\bPsi^\T \bX, y) \sim N \left\{\bH\bPsi^\T \bX + (\bPsi_0^\T - \bH\bPsi^\T)\bmu, \bD \right\},
\end{aligned}
\label{condition:LAD}
\end{equation}
where $\bv_y \in \mathbb{R}^d$ denotes a deterministic function of $y$, $\bH = (\bPsi_0^\T \bDelta \bPsi)(\bPsi^\T\bDelta \bPsi)^{-1}$ and $\bD = (\bPsi_0^\T \bDelta^{-1} \bPsi_0)^{-1}$. While condition $(i)$ allows the conditional distribution of $\bPsi^\T \bX \mid y$ to depend on $y$, condition ${(ii)}$ requires that the conditional distribution $\bPsi_0^\T \bX | (\bPsi^\T \bX, y)$ does not depend on $y$.  This aligns with the intuition that $\bPsi^\T \bX$ is a sufficient predictor for $y$. \citet{cook2009likelihood} also provide several equivalent characterisations of the two conditions in \eqref{condition:LAD}, for example, $\bPsi_0^\T \bDelta_y^{-1} = \bPsi_0^\T \bDelta^{-1}$ and $\bmu_y - \bmu = \bDelta \bPsi \bv_y$. The advantage of the conditions in equation \eqref{condition:LAD} is that a log-likelihood function can be constructed and we can maximize it to consistently estimate all the parameters of the \LAD model. As reviewed in Section 1, \citet{cook2009likelihood} highlighted that Gaussianity of the inverse predictor $\bX \mid y$ is not essential for \LAD. Indeed, their simulation results show that the \LAD estimator has superior performance to other SDR methods in terms of recovering the central subspace, even when the Gaussianity assumption is not satisfied.

When measurement errors that follow the classical additive measurement error model are present in the covariates so $\bX$ is replaced by $\bW$, condition $(ii)$ in \eqref{condition:LAD} is no longer satisfied. Specifically, 
% Based on the properties of the multivariate Gaussian distribution, 
we can straightforwardly show that the conditional mean of $\bPsi_0^\T \bW$ given $\bPsi^\T \bW$ and $y$ is
$
\text{E} ( \bPsi_0^\T \bW \vert \bPsi^\T \bW, y) = \bPsi_0^\T \bmu + \bPsi_0^\T \bDelta \bPsi \bv_y - (\bH\bPsi^\T \bDelta_y \bPsi + \bPsi_0^\T \bSigma_u \bPsi)\left\{\bPsi^\T (\bDelta_y + \bSigma_u) \bPsi\right\}^{-1}$
$(\bPsi^{\T}\bW - \bPsi^\T \bDelta\bPsi \bv_y)$.
This quantity generally depends on $y$, and so $\bPsi$ is no longer guaranteed to be a sufficient dimension reduction of the central subspace $\mathcal{S}_{y | W}$.  By a similar argument, condition $(ii)$ is also not satisfied when $\bX$ is replaced by $\hat{\bX} = \bR \bW$, with $\bR = \bSigma_x \bSigma_w^{-1}$. %The conditional mean of $\bPsi_0^\T \hat{\bX}$ given $\bPsi^\T \hat{\bX}$ and $Y=y$ is given by
%$$
%E ( \bPsi_0^\T \hat{\bX} \vert \bPsi^\T \hat{\bX}, y) = \bPsi_0^\T R\bmu + \bPsi_0^\T R\bDelta \bPsi \bv_y - (\bPsi^\T \bH_y \bPsi_0)(\bPsi^\T \bH_y \bPsi)^{-1}(\bPsi^\T R  W - \bPsi^\T R  \bDelta \bPsi \bv_y),
%$$
%where $\bH_y$ = $R (\bDelta_y + \bSigma_u) \mathbb{R}^\T$, which also depends on $y$ in general. 
We remark that this result does not contradict the invariance law of \citet{li2007surrogate}, since that was established under the assumption that the \emph{marginal} distribution of $\bX$ is Gaussian. Here, the Gaussian inverse regression model assumes only that the conditional distribution of $\bX \mid y$ is Gaussian.
% , and the marginal distribution of $\bX$ is generally not Gaussian. 

To overcome the challenges of inverse regression-based \SDR with measurement error, we propose a new predictor of $\bX$ from the observed surrogate $\bW$ which satisfies similar properties to those in \eqref{condition:LAD}, so that we can apply inverse regression based \SDR. To this end, let $\bV = \bL\bW$ with $\bL = \bDelta (\bDelta + \bSigma_u)^{-1}$. The the result below ensures that if $\bPsi$ is a semi-orthogonal basis matrix for $\mathcal{S}_{y \vert \bX}$, then it is also a semi-orthogonal basis matrix for $S_{y \vert \bV}$.
\bigskip
\begin{proposition}
Let $\bG_y = \bL(\bDelta_y + \bSigma_u) \bL^\T$ and $\tilde{\bD} = (\bPsi_0^\T \bDelta^{-1} \bL^{-1} \bPsi_0)^{-1}$. The conditional distributions $\bPsi^\T \bV | y$ and $\bPsi_0^\T \bV  \vert (\bPsi^\T \bV, y)$ are given by
\begin{gather}
\begin{aligned}
& (i) ~\bPsi^\T \bV \mid y \sim N(\bPsi^\T \bL \bmu + \bPsi^\T \bL \bDelta \bPsi \bv_y, \bPsi^\T \bG_y \bPsi), \\
& (ii) ~ \bPsi_0^\T \bV \mid (\bPsi^\T \bV, y) \sim N\{\bH\bPsi^\T \bV  + (\bPsi_0^\T - \bH\bPsi^\T) \bL\bmu, \tilde{\bD} \},
\end{aligned}   
\label{eq3}
\end{gather}

\label{proposition1}
\end{proposition}
The proof of Proposition \ref{proposition1} and all the other theoretical results can be found in Section \ref{sec:proof}. In summary, 
% similar to the case when the true covariate vector $\bX$ is observed, 
$\bV$ is purposefully constructed such that the conditional mean of $\bPsi^\T \bV \vert y$ depends on $y$, but the conditional mean of $\bPsi_0^\T \bV \vert (\bPsi^\T \bV, y)$ does not. 

\section{Maximum likelihood estimation}
\subsection{Corrected LAD estimator}
We can now consider the problem of estimating the central subspace $\mathcal{S}_{y \vert \bX}$ from the data $(y_i, \bW_i^\T)$, $i=1,\ldots, n$. Similar to \LAD and other \SDR methods based on inverse-moments, we first partition the data into $M$ non-overlapping slices based on the outcome data $y_i$, for $i=1,\ldots, n$. If the outcome is categorical, each slice corresponds to one category. If the outcome is continuous, these slices are constructed by dividing its range into $M$ non-overlapping intervals. Let $S_m \subset \{1, \ldots, n\}$ denote the index set of observations and $n_m$ be the number of observations in the $m$th slice, $m=1,\ldots,M$. Next, assume the true covariate data within each slice are mutually independent and identically distributed, such that
$
\bX_i^{(m)} \mid y_{i}^{(m)} \sim N(\bmu_{y}^{(m)}, \bDelta_{y}^{(m)})$ where 
$
\bW_i^{(m)} = \bX_i^{(m)} + \bU_i^{(m)}, ~ \bU_i^{(m)} \sim N(\bm{0}, \bSigma_u),
$
and we use the superscript $m$ to index the slice to which the $i$th observation belongs. Furthermore, we set $\text{E}\left(\bmu_y^{(m)}\right) = \text{E}\left(\bX_i^{(m)}\right) = \bmu$, and $\text{E}\left(\bDelta_y^{(m)}\right) = \bDelta$, that is, these expectations do not depend on the slice $m$. 

From Proposition \ref{proposition1}, let $\bV_{i}^{(m)} = \bL\bW_i^{(m)}$. Then we will estimate the central subspace of $\mathcal{S}_{y \vert \bX}$ by maximizing the joint log-likelihood of $\bPsi^\T \bV_i^{(m)}$ and $\bPsi_0^\T \bV_i^{(m)}$ for $i=1,\ldots, n$ and $m=1,\ldots,M$. Theorem \ref{theorem_MLE} below gives the explicit form of this log-likelihood function when the dimension $d$ is known. 

\begin{theorem}
Let $\bPsi \in \mathbb{R}^{p \times d}$ be a semi-orthogonal basis matrix for $\mathcal{S}_{y \vert \bX}$. Then the profile log-likelihood function for $\bPsi$ and $\bDelta$ from the observed data is given by
\begin{align}
\ell_1(\bPsi, \bDelta) & = -n \log \vert \bL\tilde\bSigma_w \bL^\T \vert \nonumber \\ & + n \log \vert \bPsi^\T (\bL \tilde \bSigma_w \bL^\T) \bPsi \vert  \nonumber \\ &- \sum_{m=1}^{M} n_m \log  \vert\bPsi^\T \bL \tilde\bDelta_{wm} \bL^\T \bPsi \vert, 
\label{eq:MLEjoint}
\end{align}
where $\bL = \bDelta(\bDelta + \bSigma_u)^{-1}$, $\tilde\bSigma_{w}$ and $\tilde\bDelta_{wm}$ are the sample (marginal) covariance matrix of $\bW$ and the sample covariance matrix of $\bW$ within the $m$th slice, and $\vert \bA \vert$ denotes the determinant of the square matrix $\bA$. 
\label{theorem_MLE}
\end{theorem}

The profile likelihood in Theorem 1 depends on both $\bPsi$ and $\bDelta$, so maximizing it is challenging. We propose to estimate $\bDelta$ first using a method-of-moments approach as follows: First, ignoring the measurement errors, compute the na\"ive \LAD estimator $\hat{\mathcal{S}}_{n}$  on the observed data, $(y_i, {\bW}_i^\T)$, $i=1,\ldots,n$. 
%$
%\hat{S}^{n} = \arg\min \ell_2(S) = \arg\min \{ n \log \Vert P_\mathcal{S} \tilde \bSigma  P_\mathcal{S} \Vert_0 - \sum_{m=1}^{M} n_m \log  \Vert P_\mathcal{S} \tilde\bDelta_j P_\mathcal{S} \Vert_0 \}
%$.
Let $\hat\bPsi_\text{n}$ denote a subsequent orthogonal basis of $\hat{\mathcal{S}}^{n}$. Then an estimate of $\text{E}\left\{\text{Var}(\bW \vert y) \right\}$ is given by 
$
\hat\bDelta_\text{n} = 
\{\hat\bPsi_\text{n}(\hat\bPsi_\text{n}^\T \tilde\bDelta\bPsi_\text{n})^{-1} \hat\bPsi_\text{n}^\T  + \tilde\bSigma^{-1} - \hat\bPsi_\text{n}(\hat\bPsi_\text{n}^\T \tilde\bSigma\bPsi_\text{n})^{-1} \hat\bPsi_\text{n}^\T \}^{-1}
$, and we can estimate $\hat\bDelta$ by $(\hat\bDelta_\text{n} - \bSigma_u)$. Replacing $\bDelta$ by $\hat\bDelta$ in \eqref{eq:MLEjoint} and removing terms that do not depend on $\bPsi$, we then maximize  
\begin{equation}
\begin{aligned}
\ell_2(\bPsi) & =  n \log \vert \bPsi^\T (\hat{\bL} \tilde \bSigma_{w} \hat{\bL}^\T) \bPsi \vert  \\ & - \sum_{m=1}^{M} n_m \log  \vert\bPsi^\T \hat{\bL} \tilde\bDelta_{wm} \hat{\bL}^\T \bPsi \vert, 
\end{aligned}
\label{eq:MLE2}
\end{equation}
where $ \hat{\mathbf{L}} = \bDelta(\bDelta + \bSigma_u)^{-1}$. The right-hand side of \eqref{eq:MLE2} is invariant to rotation of $\bPsi$ by an orthogonal matrix $\bA \in \mathbb{R}^{d \times d}$. That is, for any matrix $\bA$ with $\bA^\T = \bA^{-1}$, we have $\ell_2(\bA\bPsi) = \ell_2 (\bPsi)$. Therefore, the function in \eqref{eq:MLE2} is actually a function of the column space of $\bPsi$, which is characterized by the projection matrix $P_\mathcal{S} = \bPsi\bPsi^\T$. Similar to \LAD then, 
% method when the true covariate vectors $\bX_i$, $i=1,\ldots,n$, are observed, 
maximisation of \eqref{eq:MLE2} is performed over the Grassmann manifold $ \mathcal{S} \in \mathcal{G}_{d, p} \subset \mathbb{R}^{p}$, which is the subspace spanned by any $p \times d$ basis matrix. Specifically, the function in \eqref{eq:MLE2} can be written as
\begin{align}
\ell_2(\mathcal{S}) = & \log \Vert P_\mathcal{S} (\hat{\bL} \tilde \bSigma_w \hat{\bL}^\T) P_\mathcal{S} \Vert_0 \nonumber \\ & - \sum_{m=1}^{M} f_m \log  \Vert P_\mathcal{S} \hat{\bL} \tilde\bDelta_{wm} \hat{\bL}^\T P_\mathcal{S} \Vert_0, 
\label{eq: MLE_Grassmann2}
\end{align}
where for any square matrix $\bA$, we use $\Vert \bA \Vert_0$ to denote the product of its non-zero eigenvalues. % (this is also known as the pseudo-determinant). 
The corresponding Euclidean gradient of this objective function is given by    

\begin{align*}
\nabla \ell_2(\mathcal{S}) & = (\hat{\bL} \tilde\bSigma_w \hat{\bL}^\T)  \bPsi \left(\bPsi^\T \hat{\bL} \tilde\bSigma_w \hat{\bL}^\T \bPsi \right)^{-1} \\ & - \sum_{m=1}^{M} f_m  (\hat{\bL} \tilde\bDelta_{wm} \hat{\bL}^\T) \bPsi \left(\bPsi^\T \hat{\bL} \tilde\bDelta_{wm} \hat{\bL}^\T \bPsi \right)^{-1}. 
\end{align*}

\textcolor{black}{This closed-form gradient function facilitates the use of a trust region algorithm on the Grassmann manifold developed by \citet{absil2007trust}. Starting from an initial solution $\hat{\mathcal{S}}^{(0)}$, the algorithm finds the next candidate   $\hat{\mathcal{S}}^{(1)}$ by first finding a solution to minimize a constrained objective function on the tangent space at $\hat{\mathcal{S}}^{(0)}$, and then maps this solution to $\mathcal{G}_{d, p}$. The constraints imposed on the tangent space are known as a trust region. The decision to accept the candidate and to expand the region or not is based on a quotient; different values of the quotient lead to one out of three possibilities: (i) accept the candidate and expand the trust region, (ii) accept the candidate and reduce the trust region, or (iii) reject the candidate and reduce the trust region. This procedure is carried out until convergence; see \citet{gallivan2003efficient} and \citet{absil2007trust} for further details.  We use the \textsc{Manopt} package \citep{manopt} in \textsc{Matlab}, which is a dedicated toolbox for optimization on manifolds and matrices, to maximize \eqref{eq: MLE_Grassmann2}. In this implementation, there is no need to provide the Hessian of $\ell_2(\mathcal{S})$, since the \textsc{Manopt} package automatically incorporates a numerical approximation for this matrix based on finite differences in the \texttt{trustregion} solver.}

For the remainder of this article, we refer to the solution of this problem, and hence our proposed estimator, as the corrected \LAD estimator (cLAD) of the central subspace. 

\commln{
\subsection{Invariance-law \LAD estimator}
The cLAD estimator is constructed by maximizing a likelihood function involving the adjusted surrogate $\bV_i = \bL \bW_i = \bm\Delta (\bm\Delta + \bm\Sigma_u)^{-1} \bW_i$. In this section, we consider a likelihood-based estimator for the central subspace based instead on maximizing a likelihood function involving $\bX^* = \bm\Sigma_x\bm\Sigma_w^{-1}\bW$, the adjusted covariate introduced in the invariance law of \citet{li2007surrogate}. From the observed data, a sample of the adjusted covariate $\bX^*$ can be constructed by $\widehat{\bX}_i^* = \hat{\bm{\Sigma}}_x \hat{\bm\Sigma}_w^{-1}\bW_i^\top$, where $\hat{\bm\Sigma}_{w} = n^{-1}\sum_{i=1}\bW_i \bW_i^\T$ and $\hat{\bm\Sigma}_x = \hat{\bm\Sigma}_{w} -\bm{\Sigma}_u$. The equivalence between  $\mathcal{S}_{y\vert X}$ and $\mathcal{S}_{y\vert X^*}$ motivates applying \LAD to ($y_i, \hat{\bX}_i^*$), to obtain the invariance-law LAD (IL-LAD) estimator $\hat{\bm\Psi}^*$ which maximizes the objective function 
\begin{equation*}
\ell_3(\mathcal{S}) = \log \Vert P_\mathcal{S} \tilde\bSigma^* P_\mathcal{S} \Vert_0 - \sum_{m=1}^{M} f_m \log  \Vert P_\mathcal{S} \tilde\bDelta_m^* P_\mathcal{S} \Vert_0, 
\label{eq:il-lad}
\end{equation*}
where $\tilde\bSigma^*$ and $\tilde\bDelta_m^*$ denote the sample marginal covariance of $\bX^*$ and the sample covariance of $\bX^*$ within the $m$th slice, respectively. 

Similar to the cLAD estimator, maximization of \eqref{eq:il-lad} is also performed over the Grassmann manifold $ \mathcal{S} \in \mathcal{G}_{d, p}$.  The main difference between the estimators is that the cLAD estimator uses the conditional covariance $\bm \Delta = \text{E}\left\{\text{Var}(\bX \mid y\right\}$ and the IL-LAD estimator uses the marginal covariance $\bm\Sigma_x$ to construct the adjusted covariate. Nevertheless, in the following subsection, we will prove that the IL-LAD and cLAD estimator are asymptotically equivalent. In finite samples, we found that the difference between the two estimators is often negligible.

\subsection{Consistency}
In this section, we establish the consistency of both the \LAD and IL-LAD estimators, and show them to be asymptotically equivalent.  We focus on the setting when $d$ is known and $p$ is fixed, similar to \citet{cook2009likelihood}. Assuming the true covariates $\bX$ are observed, \citet{cook2009likelihood} proved the consistency of the \LAD estimator by establishing the equivalence between the population subspace spanned by \LAD and that spanned by the sliced average variance estimator i.e., the true SAVE estimator. We use a similar argument here and prove that, when $n \to \infty$, the subspace spanned by either the cLAD or IL-LAD estimators is equivalent to that of the \SAVE estimator \emph{when the true covariates $\bX$ are observed}. 

For any subspace $\mathcal{S}$, the function $n^{-1}\ell_2(\mathcal{S})$ from \eqref{eq: MLE_Grassmann2} converges to 
$
K_2(\mathcal{S}) = \log \Vert P_\mathcal{S} (\bL (\bSigma + \bSigma_u) \bL^\T) P_\mathcal{S} \Vert_0 - \sum_{m=1}^{M} f_m \log  \Vert P_\mathcal{S} {\bL} (\bDelta_y^{(m)} + \bSigma_u) \bL^\T P_\mathcal{S} \Vert_0$, where 
$
f_m = n_m/n.
$
The population cLAD subspace is then defined to be $\mathcal{S}^*_\text{LAD} = \arg \max_{S} K_2(\mathcal{S})$. Similarly, the subspace spanned by the IL-LAD estimator converges to the population IL-LAD subspace $\mathcal{S}^*_{\text{IL-LAD}}$ that maximizes $
K_3(\mathcal{S}) =  \log \Vert P_\mathcal{S} \bSigma^* P_\mathcal{S} \Vert_0 - \sum_{m=1}^{M} f_m \log  \Vert P_\mathcal{S} \bDelta_m^* P_\mathcal{S} \Vert_0,$ with $\bm\Sigma^* = \text{Var}(\bX^*) = \bm\Sigma_x\bm\Sigma_w^{-1}\bm\Sigma_x$, and $\bm\Delta_m^* = \text{Var}\left(\bX_i^{(m)*} \mid y_i^{(m)} \right) = \bm\Sigma_x \bm\Sigma_{w}^{-1}\left( \bm\Delta_y^{(m)}+ \bm\Sigma_u\right)\bm\Sigma_{w}^{-1}\bm\Sigma_x$. The main result below establishes the equivalence between the subspaces spanned by the two proposed estimators and that spanned by the true SAVE estimator. 
\begin{theorem}
$\mathcal{S}^*_\mathrm{cLAD} = \mathcal{S}^*_\mathrm{IL-LAD} = \mathcal{S}^*_\mathrm{SAVE}$, where $\mathcal{S}^*_{\mathrm{SAVE}}$ denote the population subspace spanned by the \SAVE estimator when the true covariates $\bX$ are observed.
\label{thm:consistency}
\end{theorem}

As noted in Section 3.2, the main difference between the cLAD and the IL-LAD estimators is the use of $\bm\Delta$ versus $\bm\Sigma_x$ in the construction of the adjusted surrogate. Nevertheless, asymptotically, they are both equivalent to the true SAVE estimator, since the orthogonal complement $\bm\Phi_0$ corresponding to the SAVE estimator satisfies $\bm\Phi_0^\top \bm\Delta^{-1} = \bm\Phi_0^\top \bm\Sigma_x^{-1}$.  Similar to Proposition 2 in \citet{cook2009likelihood}, Theorem \ref{thm:consistency} does not require any distributional assumptions on the model, and only depends on the properties of positive definite matrices and the concavity of the log determinant function. Nevertheless, the result implies that the cLAD estimator is consistent whenever the true \LAD estimator and \SAVE (i.e., those computed assuming $\bX$ were known) are consistent for the central subspace $\mathcal{S}_{y \vert \bX}$. As shown in \citet{li2007directional} and \citet{cook2009likelihood}, these two estimators are consistent under a linearity and constant covariance condition.

}

% , noting again that the assumption $\bX \mid y$ follows a Gaussian distribution is not required. When measurement errors $\bU$ are present, the assumption that  $\bU$ follows a Gaussian distribution is critical in establishing the properties in \eqref{proposition1}, but does not affect the relationship between the population subspaces in Theorem \ref{thm:consistency}.

\section{Sparse surrogate dimension reduction}
When the number of covariates is large, it is typically assumed that the sufficient predictors $\bB^\T \bX $ depend on only a few covariates \citep[e.g.,][]{qian2019sparse}. In other words, the true matrix $\bB$ is row-sparse. 
% , where only a few rows of $\bB$ are non-zero. 
Since only the column space of $\bB$ %that is the central subspace, 
is identifiable from the samples, we translate the sparsity of $\bB$ into the \commln{elementwise} sparsity of the projection matrix, and impose a corresponding penalty to achieve sparse solutions. \commln{Below, we formalize this idea with the proposed cLAD estimator, although an analogous procedure can be applied to the IL-LAD estimator. } 

We propose to maximize the regularized objective function
\begin{align}
\tilde{\ell}_2(\mathcal{S})  & = \ell_2(\mathcal{S}) -  \lambda \Vert P_\mathcal{S} \Vert_1 \nonumber \\ & = \log \Vert P_\mathcal{S} (\hat{\bL} \tilde \bSigma \hat{\bL}^\T) P_\mathcal{S} \Vert_0 \nonumber \\ & - \sum_{m=1}^{M} f_m \log  \Vert P_\mathcal{S} \hat{\bL} \tilde\bDelta_m \hat{\bL}^\T P_\mathcal{S} \Vert_0 - \lambda \Vert P_\mathcal{S} \Vert_1, 
\label{eq:constraintllh}
\end{align}
where for any matrix $\bA$ with elements $a_{ij}$, we denote $\Vert \bA \Vert_1 = \sum_{i,j} \vert a_{ij} \vert$, and $\lambda > 0$ is a tuning parameter. Although it is possible to use other regularization functions to achieve sparse solutions, here we follow \citet{wang2017grassmannian} and choose the $\ell_1$ norm regularizer to ease the optimization problem on the Grassmann manifold, as both the Euclidean and Riemann gradient of the objective function $\tilde{\ell}_2(\mathcal{S})$ have a closed form. 

In more detail, for any matrix $\bA$ of arbitrary dimension $m \times n$, let $\text{vec}(\bA)$ denote the $mn$-dimensional vector formed by stacking columns of $\bA$ together.
Similarly, let $\text{ivec}(\cdot)$ denote the inverse vectorization operator i.e., $\operatorname{ivec}\{(\operatorname{vec}(\bA)\} = \bA$. Let $\bT_{m,n} $ be a (unique) matrix of dimension $mn \times mn$ satisfying $\text{vec}(\bA) = \bT_{m, n} \text{vec}(\bA^\T)$. Taking the Euclidean gradient of the regularization term with respect to $\bPsi$, we obtain
\begin{align*}
&\operatorname{vec}\left\{\partial \frac{\left\|\bPsi \bPsi^{T}\right\|_{1}}{\partial\bPsi}\right\}^\T=\operatorname{vec}\left\{\operatorname{sgn}\left(\mathbf\bPsi\bPsi^\T\right)\right\}^\T \frac{\partial \bPsi \bPsi^\T}{\partial \bPsi},
\\[.5em] & \frac{\partial \bPsi\bPsi^\T}{\partial \bPsi} = \left (\bI_{p^2} + \bT_{p, p}\right)\left(\bPsi \otimes \bI_{p}\right),
\end{align*}
where $\bI_{p^2}$ is the identity matrix of dimension $p^2 \times p^2$ and $\otimes$ denotes the Kronecker product.  %For the terms in $\ell_2(S)$, the corresponding Euclidean gradient is given by
%\begin{align*}
%\nabla \ell_2(\bPsi) & = (\hat{\bL} \tilde\bSigma_w \hat{\bL}^\T)  \bPsi \left(\bPsi^\T \hat{\bL} \tilde\bSigma_w \hat{\bL}^\T \bPsi \right)^{-1} \\ & - \sum_{m=1}^{M} f_m  (\hat{\bL} \tilde\bDelta_{wm} \hat{\bL}^\T) \bPsi \left(\bPsi^\T \hat{\bL} \tilde\bDelta_{wm} \hat{\bL}^\T \bPsi \right)^{-1}. 
%\end{align*}
As a result, the Euclidean gradient of the $\tilde{\ell}_2(\mathcal{S})$ at any semi-orthogonal matrix $\bPsi$ is
$
\nabla \tilde{\ell}_2(\mathcal{S}) = \nabla \ell_2(\mathcal{S}) - \lambda \text{ivec} \left\{\text{vec}\left(\left\|\bPsi \bPsi^{T}\right\|_{1}/\partial\bPsi\right) \right\},
$
and the corresponding Riemann gradient is
$
\operatorname{grad}\tilde{\ell}_2(\bPsi) = (\bI_p - \bPsi\bPsi^\T)  \nabla \tilde{\ell}_2(\bPsi).
$
\textcolor{black}{These Euclidean and Riemann gradients are used in the same trust region algorithm of \citet{absil2007trust} to maximize $\tilde{\ell}_2(\mathcal{S})$. We computed the maximizer for \eqref{eq:constraintllh} on a grid of the tuning parameter $\lambda$ that consists of $40$ logarithmically equally-spaced values between $0$ and $\lambda_\text{max}$, where $\lambda_{\text{max}}$ is the value where the maximizer $P_{\hat{\mathcal{S}}}$ of \eqref{eq:constraintllh} is approximately close to the identity matrix. In our experiment, we found that setting $\lambda_{\max}$ to $1$ leads to a good performance and that the algorithm converges quickly to a stable solution at each value of the tuning parameter $\lambda$ in the grid.}    

Finally, to select the optimal tuning parameter $\lambda$, we use a variant of the projection information criterion (PIC) developed by \citet{nghiem2021sparse}. Let $\hat{\mathcal{S}}_0$ be an initial consistent, non-sparse estimator of the central subspace, and let $\hat{\mathcal{S}}(\lambda)$ be the maximizer of \eqref{eq:constraintllh} associated with $\lambda$. We propose to select $\lambda$ by minimizing
$
\textsc{PIC}(\lambda) = \Vert P_{\hat{\mathcal{S}}(\lambda)} - P_{\hat{S}_0}\Vert_F^2 + p^{-1}\log(p) \operatorname{df}(P_{\hat{\mathcal{S}}(\lambda)}),
$
where $\operatorname{df}(P_{\hat{\mathcal{S}}(\lambda)}) = s_\lambda(s_\lambda-d)$ with $s_{\lambda}$ being the number of non-zero, diagonal elements of $P_{\hat{\mathcal{S}}(\lambda)}$. The first term in $\textsc{PIC}(\lambda)$ is a measure of goodness-of-fit, while the second term is a measure of complexity. Compared to the information criterion developed in \citet{nghiem2021sparse}, this modified criterion has a different complexity term so as to more effectively quantify the number of parameters of the corresponding Grassmann manifold. In our numerical studies, we choose the initial consistent estimator $\hat{\mathcal{S}}_0$ to be the estimated central subspace corresponding to the unregularized cLAD estimator, and find that this choice usually leads to good variable selection performance. For the remainder of this article, we refer to the maximizer of \eqref{eq:constraintllh}, with the tuning parameter selected via this projection criterion, as the sparse corrected \LAD (scLAD) estimator.

\section{Simulation studies}
\label{section:sim}
We conduct a numerical study to examine the performance of the proposed cLAD and IL-LAD estimators in finite samples. We generate the true predictors and outcome from the following four single/multi-index models (i) $y_i = (0.5)(\bX_i^\T \bbeta_1)^3 + 0.25~ \vert \bX_i^\T \bbeta \vert \varepsilon_i$, (ii) $y_i = 3(\bX_i^\T \bbeta_1)/(1+\bX_i^\T \bbeta_1)^2 + 0.25 \varepsilon_i$, 
(iii) $y_i = 4\sin(\bX_i^\T \bbeta_2/4) + 0.5(\bX_i^\T \bbeta_1)^2 + 0.25\varepsilon_i$, and (iv) $y_i  = 3(\bX_i^\T \bbeta_1)\exp(\bX_i^\T \bbeta_2 + 0.25 \varepsilon_i) $,
with $\varepsilon_i \sim N(0,1)$, and $\bW_i = \bX_i + \bU_i$ for $i=1,\ldots, n$. The single index models (i) and (ii) are similar to those considered in \citet{Linsparse} and \citet{nghiem2021sparse}, while the multiple index models (iii) and (iv) are similar to those considered in \citet{reich2011sufficient}.  The true central subspace in models (i) and (ii) is the subspace spanned by $\bB = \bm\beta_1$, while in models (iii) and (iv) it is the column space of $\bB = [\bm\beta_1, \bm\beta_2]$. Moreover, we set $\bm\beta_1 = (1, 1, 1, 0, \ldots, 0)^\T$ and $\bm\beta_2 = (0, 0, 1, 1, 1, 0, \ldots, 0)^\T$ such that the true central subspace for the single index models depends only on the first three covariates, while for the multiple index models it depends only on the first five covariates across two indices.

Next, we generate the true predictors $\bX_i$ from one of the following three choices: (1) a $p$-variate Gaussian distribution $N(\bm{0}, \bSigma_x)$, (2) a $p$-variate $t$ distribution with three degrees of freedom and the same covariance matrix $\bSigma_x$, and (3) a $p$-variate half-Gaussian distribution $\vert N(\bm{0}, \bSigma_x) \vert$, where for all three choices we set the covariance matrix to have an autoregressive structure $\sigma_{xij} = 0.5^{\vert i-j \vert}$. Turning to the measurement error, we generate $\bU_i$ from a multivariate Gaussian distribution $N(\bm{0}, \bSigma_u)$, where $\bSigma_u$ is set to a diagonal matrix with elements drawn from a uniform $U(0.2, 0.5)$ distribution. 
% As such, the noise-to-signal ratio ranges from $20\%$ to $50\%$ on each covariate. 
The sample size $n$ is set to either $1000$ or $2000$, while the number of covariates $p$ is set to either $20$ or $40$. We assume $\bSigma_u$ and the structural dimension $d$ are known.  

It is important to highlight that the data generation processes for these simulation studies are not the same as those imposed by the Gaussian inverse regression models. Particularly, with these simulation configurations, the conditional distribution $\bX_i \mid y_i$ is generally not Gaussian due to the presence of non-linear link functions. As such, with these simulation configurations, the conditional Gaussianity only plays the role of a ``working model.'' This type of data generation process is also used in \citet{cook2009likelihood}   %Furthermore, because the invariance law of \citet{li2007surrogate} relies on the (marginal) Gaussianity of the true covariates $\bX$, we choose three distributions of $\bX$ as mentioned in the previous paragraph to examine the performance of the estimators constructed based on this law.   

For each simulated dataset, we compute the two unregularized likelihood-based estimates, cLAD and IL-LAD, and compare them with several invariance-law inverse-moment-based estimates, including \SIR (IL-SIR), the SAVE (IL-SAVE) and directional regression (IL-DR). To account for the sparsity of the central subspace, we compute the scLAD estimator and compare it with two of the invariance-law sparse estimates, namely the Lasso \SIR of \citet{Linsparse} (IL-Lin) and the sparse \SIR proposed by \citet{tan2018convex} (IL-Tan). Both of these estimators are sparse versions of sliced inverse regression, where the first imposes sparsity on each sufficient direction and the second imposes sparsity on the projection matrix and induces row-sparsity. For the estimator proposed by \citet{Linsparse}, we use the \textsc{R} package \textsc{LassoSIR} with its default settings. For the estimator proposed by \citet{tan2018convex}, we use the code provided by the authors.
% and implement it with the default settings. 
For both estimators, we follow the recommend procedure from the authors and choose the tuning parameter based on a ten-fold cross-validation procedure. 
% The tuning parameter for scLAD is selected based on the proposed projection information criterion. 

\commln{We assess the performance of all estimators based on the Frobenius norm of the difference between the projection matrix associated with the true central subspace, and that of the estimated central subspace. For the sparse estimators, we assess variable selection performance based on the following metric. Let $\mathbf{P}$ denote the projection matrix associated with the true directions, i.e., $\mathbf{P} = \bB\left(\bB^\T \bB \right)^{-1}\bB^\T$, and let $\hat{\mathbf{P}}$ be an estimator of $\mathbf{P}$. We then define the number of true positives (TP) to be the number of non-zero diagonal elements that are correctly identified to be non-zeros,
% i.e $\text{TP} = \sum_{j=1}^{p} 1(p_{jj} \neq 0) 1(\hat{p}_{jj} \neq 0)$,  
the number of false negatives (FN) to be the number of non-zero diagonal elements that are incorrectly identified to be zeros
% , i.e $\text{FN} = \sum_{j=1}^{p} 1(p_{jj}\neq 0) 1(\hat{p}_{jj} = 0)$, 
and the number of false positives (FP) to be the number of zero diagonal elements that are incorrectly identified to be non-zeros.
% , i.e $\text{FP} = \sum_{j=1}^{p} 1(p_{jj}= 0) 1(\hat{p}_{jj} \neq  0)$, where $1(\cdot)$ is the indicator function. 
We also compute the F1 score as {2TP}/(2TP + FP + FN). This metric ranges from zero to one, with a value of one indicating perfect variable selection.
% and a higher value of F1 score is desirable. 

For brevity, we present the projection error and F1 variable selection results for $p=40$ below; the results for $p=20$ and including FP and FN selection results offer overall similar conclusions to those seen below and are deferred to the Supplementary Materials. Table \ref{tab: unregularizedestimates_p40} demonstrates that among the unregularized estimators, the likelihood-based estimators cLAD and IL-LAD have smaller estimation error in the majority of settings compared to the other, invariance-law inverse-moment-based estimators.  This result is consistent with \citet{cook2009likelihood} who demonstrated that \LAD in general tends to exhibit superior performance relative to other inverse-moment-based estimators like SIR and SAVE when no measurement error is present. The performance of all the considered estimators tends to deteriorate when the true covariates $\bX$ deviate from Gaussianity, such as when they are skewed or have heavier tails. There is negligible difference in the performance between the cLAD and IL-LAD estimator. Next, the left half of Table \ref{tabs:sim_regularizedestimates_p40} demonstrates the estimation performance of the sparse \SDR estimators. In terms of projection error, and analogous to the results with unregularized methods above, the proposed scLAD estimator 
% with the tuning parameter selected via the projection information criterion 
tends to have lower estimation error than the IL-Lin and IL-Tan estimators. The improvement is most pronounced in the two multiple index models, i.e.\ Model (iii) and Model (iv), and when the true covariate-vector $\bX$ follows a Gaussian distribution. Again similar to the unregularized estimators, the performance of these sparse estimators deteriorates when the true covariates $\bX$ deviate from Gaussianity, especially when they follow a heavy tail distribution such as $t_3$. 
% latex table generated in R 4.1.2 by xtable 1.8-4 package
% Sat Jul 30 14:24:00 2022
\begin{table*}[ht]
\centering
%\begin{threeparttable}
\caption{Mean projection errors of unregularized estimators in simulation settings with $p=40$. The lowest value in each row is highlighted. }   
\begin{tabular}{llrrrrrr}
\toprule[1.5pt]
$\bX$ & Model & $n$   & cLAD & IL-LAD&  IL-SIR & IL-SAVE & IL-DR\\ 
  \hline
  \addlinespace
$N(\boldsymbol{0}, \bm\Sigma_x)$  & (i) & 1000 & \bf{0.19} & \bf{0.19} & 0.20 & 0.32 & 0.27 \\ 
   &  & 2000 & \bf{0.13} & \bf{0.13} & 0.14 & 0.16 & 0.19 \\ 
   & (ii) & 1000 & \bf{0.29} & \bf{0.29} & 0.42 & 1.01 & 0.43 \\ 
   &  & 2000 & \bf{0.20} & \bf{0.20} & 0.29 & 0.37 & 0.29 \\ 
   & (iii) & 1000 & \bf{0.69} & \bf{0.69} & 1.40 & 1.55 & 0.95 \\ 
   &  & 2000 & \bf{0.48} & \bf{0.48} & 1.20 & 0.96 & 0.68 \\ 
   & (iv) & 1000 & \bf{0.42} & 0.43 & 0.54 & 1.53 & 0.61 \\ 
   &  & 2000 & \bf{0.29} & \bf{0.29}
   & 0.38 & 0.58 & 0.42 \\[.5em]
   
   $\vert N(\boldsymbol{0}, \bm\Sigma_x) \vert $ & (i) & 1000 & \bf{0.23} & \bf{0.23} & 0.24 & 0.96 & 0.31 \\ 
   &  & 2000 & \bf{0.16} & \bf{0.16} & 0.17 & 0.24 & 0.22 \\ 
   & (ii) & 1000 & \bf{0.41} & \bf{0.41} & 0.46 & 1.40 & 0.56 \\ 
   &  & 2000 & \bf{0.29} & \bf{0.29} & 0.33 & 1.13 & 0.41 \\ 
   & (iii) & 1000 & \bf{1.06} & 1.09 & 1.28 & 1.47 & 1.24 \\ 
   &  & 2000 & \bf{0.78} & \bf{0.78} & 1.10 & 1.20 & 1.05 \\ 
   & (iv) & 1000 & \bf{1.30} & \bf{1.30} & 1.37 & 1.49 & 1.35 \\ 
   &  & 2000 & \bf{1.15} & \bf{1.15} & 1.28 & 1.25 & 1.22 \\[.5em] 
  
  $t_3$ & (i) & 1000 & \bf{0.37} & \bf{0.37} & 0.55 & 1.41 & 1.17 \\ 
   &  & 2000 & \bf{0.28} & \bf{0.28} & 0.48 & 1.41 & 1.06 \\ 
   & (ii) & 1000 & \bf{0.69} & 0.71 & 0.77 & 1.40 & 1.39 \\ 
   &  & 2000 & \bf{0.51} & 0.52 & 0.63 & 1.40 & 1.38 \\ 
   & (iii) & 1000 & \bf{1.57} & \bf{1.57} & 1.63 & 1.86 & 1.84 \\ 
   &  & 2000 & \bf{1.41} & 1.42 & 1.54 & 1.80 & 1.77 \\ 
   & (iv) & 1000 & 1.21 & 1.21 & \bf{0.96} & 1.98 & 1.80 \\ 
   &  & 2000 & 1.02 & \bf{1.01} & 0.80 & 1.98 & 1.69 \\ 
   \bottomrule[1.5pt]
\end{tabular}
\label{tab: unregularizedestimates_p40}
%\end{threeparttable}
\end{table*}

% latex table generated in R 4.1.2 by xtable 1.8-4 package
% Sat Jul 30 16:10:50 2022
\begin{table*}[htb]
\centering
 %\begin{threeparttable}
\caption{Performance of the sparse estimators based on  average projection error and  F1 score for simulation settings with $p=40$. The lowest projection error and the highest F1 score in each row are highlighted.}
\begin{tabular}{lllrrrrrr}
\toprule[1.5pt]
$\bX$ & Model & $n$ & \multicolumn{3}{c}{Projection error} & \multicolumn{3}{c}{F1} \\  
\cmidrule(lr){4-6} \cmidrule(lr){7-9}
& & & scLAD & IL-Lin & IL-Tan & scLAD & IL-Lin & IL-Tan  \\ 
  \hline
  \addlinespace
$N(\boldsymbol{0}, \bm\Sigma_x)$  & (i) & 1000 & \bf{0.05} & 0.09 & 0.13 & \bfit{1.00} & 0.58 & \bfit{1.00} \\ 
   &  & 2000 & \bf{0.04} & 0.08 & 0.13 & \bfit{1.00} & 0.60 & 0.99 \\ 
   & (ii) & 1000 & \bf{0.09} & 0.15 & 0.19 & \bfit{1.00} & 0.61 & 0.96 \\ 
   &  & 2000 & \bf{0.05} & 0.11 & 0.14 & \bfit{1.00} & 0.64 & 0.99 \\ 
   & (iii) & 1000 & \bf{0.30} & 1.04 & 1.10 & \bfit{0.99} & 0.41 & 0.89 \\ 
   &  & 2000 & \bf{0.19} & 0.85 & 1.10 & \bfit{1.00} & 0.43 & 0.96 \\ 
   & (iv) & 1000 & \bf{0.17} & 0.71 & 0.77 & \bfit{1.00} & 0.42 & 0.95 \\ 
   &  & 2000 & \bf{0.10} & 0.55 & 0.72 & \bfit{1.00} & 0.41 & 0.98 \\[.5em]
   
  $\vert N(\boldsymbol{0}, \bm\Sigma_x) \vert $ & (i) & 1000 & \bf{0.07} & 0.11 & 0.16 & \bfit{1.00} & 0.49 & \bfit{1.00} \\ 
   &  & 2000 & \bf{0.05} & 0.09 & 0.16 & \bfit{1.00} & 0.46 & \bfit{1.00} \\ 
   & (ii) & 1000 & \bf{0.16} & 0.17 & 0.28 & \bfit{1.00} & 0.57 & \bfit{1.00} \\ 
   &  & 2000 & \bf{0.11} & 0.13 & 0.26 & \bfit{1.00} & 0.56 & \bfit{1.00} \\ 
   & (iii) & 1000 & \bf{0.97} & 1.06 & 1.21 & 0.84 & 0.39 & \bfit{0.93} \\ 
   &  & 2000 & \bf{0.72} & 0.85 & 1.18 & \bfit{0.99 }& 0.41 & 0.93 \\ 
   & (iv) & 1000 & \bf{1.24} & 1.38 & 1.34 & 0.78 & 0.36 & \bfit{0.89} \\ 
   &  & 2000 & \bf{1.11} & 1.39 & 1.36 & \bfit{0.96} & 0.34 & 0.89 \\[.5em]
   
  $t_3$ & (i) & 1000 & \bf{0.19} & 0.30 & 0.37 & \bfit{0.94} & 0.51 & 0.89 \\ 
   &  & 2000 & \bf{0.15} & 0.29 & 0.36 & \bfit{0.95} & 0.52 & 0.89 \\ 
& (ii) & 1000 & {0.51} & \bf{0.37} & 0.39 & 0.75 & 0.56 & \bfit{0.82} \\ 
   &  & 2000 & 0.34 & 0.33 & \bf{0.32} & 0.84 & 0.54 & \bfit{0.88} \\ 
   & (iii) & 1000 & \bf{1.39} & 1.44 & 1.61 & \bfit{0.55} & 0.46 & 0.50 \\ 
   &  & 2000 & \bf{1.22} & 1.35 & 1.64 & \bfit{0.61} & 0.45 & 0.49 \\ 
   & (iv) & 1000 & 1.02 & \bf{1.01} & 1.28 & \bfit{0.72} & 0.44 & 0.69 \\ 
   &  & 2000 & \bf{0.83} & 0.84 & 1.31 & \bfit{0.78} & 0.45 & 0.73 \\ 
   \bottomrule[1.5pt]
\end{tabular}
\label{tabs:sim_regularizedestimates_p40}
%\end{threeparttable}
\end{table*}   

\FloatBarrier

Turning to the variable selection results, the scLAD estimator had the best overall selection performance among the three considered estimators. When the true covariate vector $\bX$ follows a Gaussian distribution, the scLAD estimator selects the true set of important covariates across all considered settings, as reflected in the corresponding F1 scores all being exactly equal to one. 
% The IL-Tan estimator has variable selection performance close to the scLAD estimator. 

This conclusion still holds when the true covariates follow a half-Gaussian distribution and under the single index models (i) and (ii). However, for the other settings such as the multiple index models (iii) and (iv), scLAD tends to have lower F1 scores than IL-Tan for $n=1000$, although the trend is reversed when the sample size increases to $n=2000$. By contrast, the IL-Lin estimator consistently has a very low F1 score, with additional results in the Supplementary Materials demonstrating that this estimator incurs too many false positives. This reflects the advantage of imposing regularisation directly on the diagonal elements of the corresponding projection matrix, compared to doing so on each dimension separately. However, we acknowledge that the variable selection results do depend on how the tuning parameter is chosen, and we are not aware of any method that is guaranteed to achieve selection consistency when we have to estimate the central subspace from surrogates.}

% In summary, the numerical results demonstrate that the proposed cLAD and scLAD estimators have strong performance to the estimators constructed based on the invariance law of \citet{li2007surrogate}.
% Our simulations also show that when the true covariates are non-Gaussian distributed, the performance of the estimators can be affected greatly. 

\section{National Health and Nutrition Examination Survey data}
We apply the proposed methodology to analyze a dataset from the National Health and Nutrition Examination Survey (NHANES) \citep{nhanesdata}. This survey aims to assess the health and nutritional status of people in the United States and track the evolution of this status over time. During the 2009-2010 survey period, participants were interviewed and asked to provide their demographic background as well as information about nutrition habits, and to undertake a series of health examinations. To assess the nutritional habits of participants, dietary data were collected using two 24-hour recall interviews wherein the participants self-reported the consumed amounts of a set of food items during the 24 hours prior to each interview. Based on these recalls, daily aggregated consumption of water, food energy, and other nutrition components such as total fat and total sugar consumption were computed. 

In this application, we focus on the relationship between participants' total cholesterol level ($y$), their age ($Z$), and their daily intakes of $42$ nutrition components, such as sugars, total vitamins, fats, retinol, lycopene,  zinc, and selenium, among many others. We restrict our analysis to $n = 3343$ women, and assume participants' ages are measured accurately while the daily intakes of nutrition components are subject to additive measurement errors. For the $i$th participant, let $\bW_{i1}$ and $\bW_{i2}$ denote the $43 \times 1$ vector of surrogates at the first and second interview, respectively.  For each vector,  the first element is the age ($Z_i$) and the remaining elements are the recorded values for nutrition components at the corresponding interview time. We assume the classical measurement error model $\bW_{ij} = \bX_i + \bU_{ij}$ for $i=1,\ldots,n$ and $j=1,2$, where $\bX_i$ denotes the vector of the long--term nutrition intakes, and $\bU_i \sim N(\bm{0}, {\bSigma}^*_u)$ denotes the measurement errors. Moreover, as $\text{E}\left\{(\bW_{i1}- \bW_{i2}) (\bW_{i1}- \bW_{i2})^\T \right\} = 2{\bSigma}^*_u$, we estimate $\widehat{{\bSigma}}^*_u = (2n)^{-1} \sum_{i=1}^{n} (\bW_{i1}- \bW_{i2}) (\bW_{i1}- \bW_{i2})^\T$. Therefore, the covariance matrix of the measurement errors corresponding to $\bW_i = (1/2)(\bW_{i1} + \bW_{i2})$ is estimated as $\hat\bSigma_u = \widehat{\bSigma}_u^*/2$. Note since the age is assumed to be measured without error, the first row and column of $\hat\bSigma_u$ are zero.

We estimate the central subspace $\mathcal{S}_{y \vert \bX}$ using the three sparse estimators in the simulation studies, setting $d=1$ and $H=20$ slices. Letting $\hat\bbeta^{(\text{IL-Tan})}$  and  $\hat\bbeta^{(\text{IL-Lin})}$ be the estimated bases from the IL-Tan and IL-Lin estimators, respectively, the sufficient predictors are estimated to be $\hat{T}_i^{(\text{IL-Tan})} = \hat{\bX}_i\hat\bbeta^{(\text{IL-Tan})}$ and $\hat{T}_i^{(\text{IL-Lin})}=\hat{\bX}_i\hat\bbeta^{(\text{IL-Lin})}$, where $\hat{\bX}_i = (\hat\bSigma_w-\hat\bSigma_u)\hat\bSigma_w^{-1} \bW_i$. 
For the scLAD estimator $\hat{\bm\beta}^{(\text{scLAD})}$, we denote the corresponding sufficient predictor as $ \hat{T}_i^{(\text{scLAD})} = \hat{\bV}_i\hat\bbeta^{(\text{scLAD})}$, where $\hat{\bV}_i = \hat{\bDelta}(\hat\bDelta + \hat\bSigma_u)^{-1}\bW_i$. Figure 1 presents the plots of $y$ against the three sufficient predictors. 

\begin{figure*}[htb]
\centering
\includegraphics[width = \textwidth]{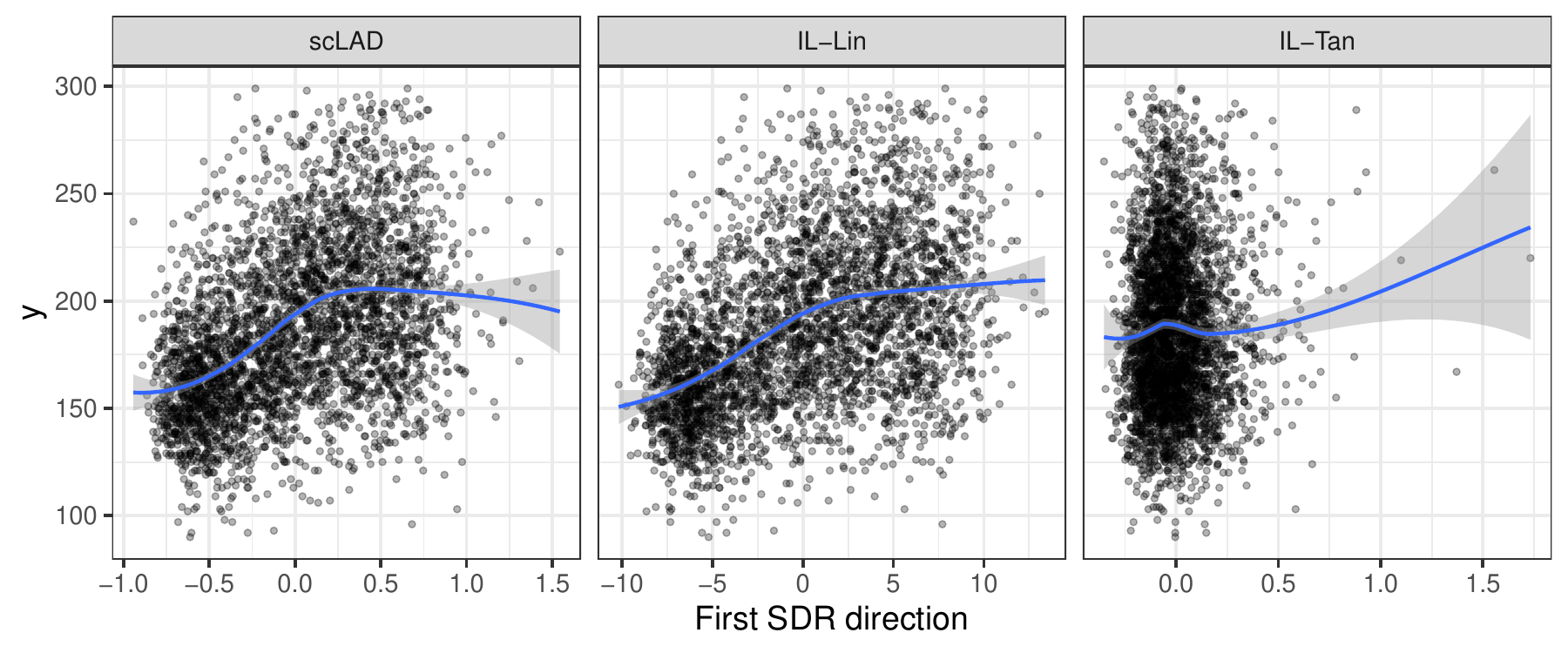}
\caption{Scatterplots of the outcome versus sufficient predictors obtained from the three sparse estimators of the central subspace in the application to the NHANES data.  The blue curves are LOESS smoothing curves and the grey areas are confidence bands.}
\end{figure*}

It can be seen that the sufficient predictors formed from scLAD and IL-Lin estimators are relatively similar to each other, and are more informative about the total cholesterol level than the IL-Tan estimator. Indeed, the sample correlation between the outcome and $\hat{\mathbf{T}}^{(\text{IL-Tan})}$ is 0.04, but the sample correlations between the outcome and $\hat{\mathbf{T}}^{(\text{IL-Lin})}$ and $\hat{\mathbf{T}}^{(\text{IL-scLAD})}$ are 0.45 and 0.43, respectively. 
In terms of variable selection, out of 43 predictors, the IL-Lin estimator selects 27 variables, the IL-Tan estimator selects 15 variables, while the proposed scLAD estimator selects 17 variables. The simulation results suggest that IL-Lin is potentially overfitting. Among the variables selected by scLAD, the three predictors that have the largest magnitudes are copper, vitamin B12, and zinc.  

\section{Discussion}
In this article, we propose two likelihood-based estimators for the central subspace when the predictors are contaminated by additive measurement errors. These estimators are constructed from a Gaussian inverse regression model $\bX \mid y$ into which measurement error is incorporated in a straightforward manner. The two estimators are based on maximizing the likelihood-based objective functions of two different adjusted covariates, the first one using the conditional covariance $\text{E}\left\{ \text{Var}(\bX \mid y) \right\}$ and the second one using the marginal covariance of $\mathbf{X}$. We establish the asymptotic equivalence of these two estimators. When the number of covariates is large, we propose a sparse corrected \LAD estimator that facilitates variable selection by introducing a penalty term that regularizes the projection matrix directly. The corresponding objective function is defined on an appropriate Grassmann manifold, and has closed-form gradients that facilitate efficient computation. Simulation studies generating data from forward index models demonstrate that the likelihood-based estimators have superior performance to the inverse moment-based estimators in terms of both estimation and variable selection consistency. Note that in our simulation studies, we did not consider generating data directly from the inverse regression model; we anticipate that the overall conclusions would be similar to, if not better than in terms of favoring our proposed estimators, those presented in this article with forward index models.

Future research can address the theoretical properties of the sparse likelihood-based estimators, particularly their variable selection consistency, and develop  likelihood-based estimators of the central subspace that are robust to non-Gaussian measurement errors.  Furthermore, the issue of estimating the number of dimensions $d$ from the surrogate data remains open. 

\appendix
%\section{Appendix}
%\label{sec:proof}
\subsection*{Appendix A: Proof of Proposition 1}
Throughout the proof, we use the following equalities from \citet[p. 205] {cook2009likelihood}, which are from \citet[p. 77]{rao1973linear}.  Let $\bB \in \mathbb{R}^{p\times p}$ be a symmetric positive definite matrix, and $(\bPsi, \bPsi_0) \in \mathbb{R}^{p \times p}$ be an orthogonal matrix. Then we have
\begin{equation}
(\bPsi^\T \bB \bPsi_0)(\bPsi^\T \bB \bPsi)^{-1} = - (\bPsi_0^\T \bB^{-1} \bPsi_0)^{-1} (\bPsi_0^\T \bB^{-1}\bPsi) \label{eq: matrixequality}
\end{equation}
\begin{equation}
(\bPsi_0^\T \bB^{-1} \bPsi_0)^{-1} = \bPsi_0^\T \bB \bPsi_0 - (\bPsi^\T \bB \bPsi_0) (\bPsi^\T \bB \bPsi)^{-1} (\bPsi_0^\T \bB \bPsi)\label{eq:matrix2}
\end{equation}
\begin{equation}
\bPsi(\bPsi^\T \bB \bPsi)^{-1}\bPsi^\top  + \bB^{-1}\bPsi_0 (\bPsi_0^\T \bB^{-1}\bPsi_0)^{-1}\bPsi_0^\T \bB^{-1} = \bB^{-1}. \label{eq:matrix3} 
\end{equation}
Recall that $\bV = \bL\bW$ with $\bL = \bDelta(\bDelta + \bSigma_u)^{-1}$, and the inverse Gaussian model $\bX \mid y \sim N(\bmu_y, \bDelta_y)$ with $\bmu_y - \bmu = \bDelta \bPsi \bv_y$ and $\bPsi_0 \bDelta_y^{-1} = \bPsi_0 \bDelta^{-1}$. 
The joint distribution of $\bPsi^\T \bV$ and $\bPsi_0^\T \bV$ given $y$ is
\begin{equation}
    (\bPsi^\T \bV,
    \bPsi_0^\T \bV) \mid y  \sim N \left(\bmu^{**}_j, \tilde{\bSigma}^{**} \right),
\end{equation}
where 
\begin{align*}
&\bmu^{**}_j = \begin{pmatrix}
\bPsi^\T \bL\bmu + \bPsi^\T \bL\bDelta \bPsi \bv_y \\
\bPsi_0^\T \bL\bmu + \bPsi_0^\T \bL\bDelta \bPsi \bv_y \\
\end{pmatrix}, \\ & \tilde{\bSigma}^{**}_j = \begin{bmatrix}
\bPsi^\T  \bG_y \bPsi & \bPsi_0^\T \bG_y \bPsi \\
\bPsi^\T \bG_y \bPsi_0 & \bPsi_0^\T \bG_y \bPsi_0 \\
\end{bmatrix}
\end{align*}
with $\bG_y = \bL(\bDelta_y + \bSigma_u) \bL^\T = \bDelta (\bDelta + \bSigma_u)^{-1} (\bDelta_y + \bSigma_u) (\bDelta + \bSigma_u)^{-1} \bDelta$. Hence part (i) of Proposition 1 follows. For part (ii), the conditional distribution $\bPsi_0^\T \bV \mid (\bPsi^\T \bV, y)$ is normal. The conditional mean is
\begin{equation}
\begin{aligned}
& \text{E}\left(\bPsi_0^\T \bV \vert \bPsi^\T \bV, y \right) =  \bPsi_0^\T \bL\bmu + \bPsi_0^\T \bL\bDelta \bPsi \bv_y  \\ & - (\bPsi^\T \bG_y \bPsi_0)(\bPsi^\T \bG_y \bPsi)^{-1}(\bPsi^\T \bL\bW - \bPsi^\T \bL \bmu - \bPsi^\T \bL\bDelta\bPsi \bv_y)
\end{aligned}
\label{eq:eq5}
\end{equation}
This conditional mean does not depend on $y$; indeed, the coefficient associated with $\bv_y$ equals
\begin{equation}
\bPsi_0^\T \bL\bDelta \bPsi  - (\bPsi^\T \bG_y \bPsi_0)(\bPsi^\T \bG_y \bPsi)^{-1}(\bPsi^\T \bL \bDelta\bPsi) = 0.
\label{eq:6}
\end{equation}
To see the last equality, applying \eqref{eq: matrixequality} with $\bB = \bG_y$, we have
$
(\bPsi^\T \bG_y \bPsi_0)(\bPsi^\T \bG_y \bPsi)^{-1} = - (\bPsi_0^\T \bG_y^{-1} \bPsi_0)^{-1} (\bPsi_0^\T \bG_y^{-1}\bPsi),
$
and furthermore, 
\begin{equation*}
\begin{aligned}
& \bPsi_0^\T \bG_y ^{-1} \\ & = \bPsi_0^\T \bDelta^{-1} (\bDelta + \bSigma_u) (\bDelta_y + \bSigma_u)^{-1}(\bDelta + \bSigma_u) \bDelta^{-1} \\
& = (\bPsi_0^\T + \bPsi_0^\T \bDelta^{-1}\bSigma_u)(\bDelta_y + \bSigma_u)^{-1}(\bDelta + \bSigma_u) \bDelta^{-1}\\ 
& \stackrel{(a)}{=} (\bPsi_0^\T + \bPsi_0^\T \bDelta_y^{-1}\bSigma_u) (\bDelta_y + \bSigma_u)^{-1}(\bDelta + \bSigma_u) \bDelta^{-1} \\
& = \bPsi_0^\T \bDelta_y^{-1}( \bDelta_y + \bSigma_u)(\bDelta_y + \bSigma_u)^{-1}(\bDelta + \bSigma_u)\bDelta^{-1} \\
& \stackrel{(b)}{=} \bPsi_0^\T \bDelta^{-1}\bL^{-1},
\end{aligned}
\end{equation*}
where steps $(a)$ and $(b)$ follow from $\bPsi_0^\T \bDelta_y^{-1} = \bPsi_0^\T \bDelta^{-1}$.  Therefore, 
$$
\begin{aligned}
\bH & \coloneqq (\bPsi^\T \bG_y \bPsi_0)(\bPsi^\T \bG_y \bPsi)^{-1} \\ & = - (\bPsi_0^\T \bG_y^{-1} \bPsi_0)^{-1} (\bPsi_0^\T \bG_y^{-1}\bPsi) \\ & = - (\bPsi_0^\T \bDelta^{-1} \bL^{-1}\bPsi_0)^{-1}(\bPsi_0^\T \bDelta^{-1} \bL^{-1} \bPsi) \\ & = (\bPsi_0^\T \bL\bDelta \bPsi)(\bPsi^\T \bL\bDelta \bPsi)^{-1},
\end{aligned}
$$
where the last equality follows from applying \eqref{eq: matrixequality} with $\bB = \bL\bDelta$.
Substituting it into the left hand side of \eqref{eq:6} gives the claim. Hence the conditional expectation in \eqref{eq:eq5} is reduced to
$
\text{E}\left(\bPsi_0^\T \bV \vert \bPsi^\T \bV, y \right)  = \bPsi_0^\T \bL\bmu +  (\bPsi_0^\T \bL\bDelta \bPsi)(\bPsi^\T \bL\bDelta \bPsi)^{-1}(\bPsi^\T \bL\bW - \bPsi^\T \bL\bmu) = (\bPsi_0^\T - \bH\bPsi^\T) \bL \bmu + \bH\bPsi^\T \bL\bW.
$
Next, the conditional covariance matrix of $\bPsi_0^\T \bV $ given $\bPsi^\T \bV$ and $y$ equals  
\begin{align*}
\tilde{\bD} & \coloneqq  \bPsi_0^\T \bG_y \bPsi_0 - (\bPsi^\T \bG_y \bPsi_0) (\bPsi^\T \bG_y \bPsi)^{-1} (\bPsi_0^\T \bG_y \bPsi)  \\ & = (\bPsi_0^\T \bG_y^{-1} \bPsi_0)^{-1} = (\bPsi_0^\T \bDelta^{-1}\bL^{-1} \bPsi_0)^{-1},
\end{align*}
which does not depend on $y$ either, where the second equality follows from \eqref{eq:matrix2} with $\bB = \bG_y$. The proof is hence complete. 

\subsection*{Appendix B: Proof of Theorem 1}
We estimate the central subspace of $\mathcal{S}_{y \vert \bX}$ by maximizing the joint log likelihood of $\bPsi^\T \bV_i^{(m)}$ and $\bPsi_0^\T \bV_i^{(m)}$ for $i=1,\ldots, n$ and $m=1,\ldots, M$. Let $n_m$ denote the number of observations in the $j$th slice, and $f_m = n_m/n$. Let $\overline{\bW}_m$ and $\tilde{\bDelta}_m$ denote the sample mean and sample covariance of the surrogate data within the $m$th slice, respectively. Let $\bG_m = \bL(\bDelta_y^{(m)} + \bSigma_u) \bL^\T = \bDelta (\bDelta + \bSigma_u)^{-1} (\bDelta_y^{(m)} + \bSigma_u) (\bDelta + \bSigma_u)^{-1} \bDelta$. For any square matrix $\bA$, we use $\operatorname{tr}(\bA)$ to denote its trace. Then,  the joint log-likelihood of the data is given by 
\begin{equation}
\begin{aligned}
2 \ell_d & = - \sum_{m=1}^{M} n_m \log \vert \bPsi^\T \bG_m \bPsi \vert - n \log \vert \bD \vert \\
& - \sum_{m=1}^{M} n_m \left(\bPsi^\T \bL \overline{\bW}_m - \bPsi^\T \bL \bmu - \bPsi^\T \bL \bDelta \bPsi \bv_m \right)^{\T} (\bPsi^\T \bG_m \bPsi)^{-1} \\ & \times \left(\bPsi^\T \bL \overline{\bW}_m - \bPsi^\T \bL \bmu - \bPsi^\T \bL \bDelta \bPsi \bv_m \right) \\ 
& - \sum_{m=1}^{M} n_m \left\{\bK^\T \bL \left(\overline{\bW}_m - \bmu \right) \right\}^{\T} \bD^{-1} \left\{\bK^\T \bL (\overline{\bW}_m - \bmu) \right\}  \\ & - \sum_{m=1}^{M} n_m \text{tr}\left\{(\bPsi^\T \bL \tilde{\bDelta}_m \bL^\T \bPsi \left(\bPsi^\T \bG_m \bPsi \right)^{-1}\right\} \\ 
& - \sum_{m=1}^{M} n_m \text{tr}\left(\bL^\T \bK \bD^{-1} \bK^\T \bL \tilde{\bDelta}_m \right), 
\end{aligned}
\end{equation}
with $ \bK \coloneqq \bPsi_0 - \bPsi \bH^\T$. We maximize the log likelihood with the two constraints that $\sum_{m=1}^{M} f_m \bv_m = 0$, and $\bPsi^\T \left( \sum_{m=1}^{M} f_m \bG_m \right) \bPsi = \bPsi^\T \bL\bDelta\bPsi $. First, we maximize with respect to $\bv_m$; only the third term in the log-likelihood above contains $\bv_m$ so combining this term with the first constraint, we need to find $\bv_m$ that minimizes
\begin{equation}
\sum_{m=1}^{M} f_m (\bZ_m - \bB \bv_m)^\T \bB_m^{-1} (\bZ_m - \bB \bv_m) + \lambda \sum_{i=1} f_m \bv_m,
\label{eq: bv_m}
\end{equation}
where $\bZ_m \coloneqq \bPsi^\T \bL(\overline{\bW}_m - \bmu)$, $\bB \coloneqq \bPsi^\T \bL\bDelta\bPsi$, and $\bB_m \coloneqq \bPsi^\T \bG_m \bPsi$. Differentiating \eqref{eq: bv_m} with respect to $\bv_m$ and setting the derivatives equal zero, we obtain  $2 f_m\bB\bB_m^{-1}(\bZ_m - \bB \bv_m) + \lambda f_m  = 0$, or equivalently, $2f_m \bZ_m - \bB f_m\bv_m +  f_m\bB_m\bB^{-1}\lambda = 0 $. Summing over all $m$ and using the constraints, we have $\lambda = \sum_{m=1}^{M} f_m \bZ_m \coloneqq \bar{\bZ}$, and hence $\bZ_m - \bB \bv_m = \bB_m \bB^{-1} \lambda $. Therefore, at the optimized value of $\bv_m$, \eqref{eq: bv_m} reduces to
$$
\begin{aligned}
&\sum_{m=1}^{M} f_m(\bB_m \bB^{-1} \lambda)^\T \bB_m^{-1}(\bB_m \bB^{-1} \lambda)  \\ & = \lambda^\T \bB^{-1} \left(\sum_{m=1}^{M} f_m \bB_m \right) \bB^{-1} \lambda  \\ & = \lambda^\T \bB^{-1} \bB \bB^{-1} \lambda = \lambda^\T \bB^{-1} \lambda = \bar{\bZ}^\T \bB^{-1}\bar{\bZ}\\&
= \left\{\lambda^\T \bL(\overline{\bW} - \bmu) \right\}^\T \bB^{-1} \left\{\lambda^\T \bL(\overline{\bW} - \bmu) \right\}, 
\end{aligned}
$$
with $\overline{\bW} = \sum_{m=1}^{M} f_m \overline{\bW}_m$. 

Next, we maximize the log-likelihood with respect to $\bmu$ at the optimized value of $\bv_m$. Taking the derivative with respect to $\bmu$ and setting it equal to zero, we obtain
\begin{align}
& \bL^\T \bPsi \bB^{-1} \bPsi^\T \bL (\overline{\bW}-\bmu) + \bL^\T \bK \bD^{-1} \bK^\T \bL (\overline{\bW}-\bmu) \nonumber  \\ & = \bL^\T (\bPsi \bB^{-1} \bPsi^\T + \bK \bD^{-1} \bK^\T) \bL(\overline{\bW}-\bmu) = 0. 
\label{eq:mu}
\end{align}
Noting that 
\begin{align*}
\bK^\T & = \bPsi_0^\T - \bH\bPsi^\T  \\ & = \bPsi_0^\T - \bPsi_0^\T \bL\bDelta \bPsi (\bPsi^\T \bL \bDelta \bPsi)^{-1} \bPsi^\T \\ 
& = \bPsi_0^\T (\bI_p - (\bL\bDelta)\bPsi(\bPsi^\T \bL\bDelta\bPsi)^{-1}\bPsi^\T)\\
%& = \bPsi_0^\T \left(\bI_p - P^\T_{\bPsi(\bL\bDelta)} \right) \\
%& = \bPsi_0^\T P_{{\bPsi_0}(\bL\bDelta)^{-1}} \\ 
& = \bPsi_0^\T \bPsi_0 (\bPsi_0^\T \bDelta^{-1} \bL^{-1} \bPsi_0)^{-1}\bPsi_0^\T (\bL\bDelta)^{-1} \\ 
& =(\bPsi_0^\T \bDelta^{-1} \bL^{-1} \bPsi_0)^{-1}\bPsi_0^\T \bDelta^{-1}\bL^{-1}  \\ &= \bD \bPsi_0^\T (\bL\bDelta)^{-1}, 
\end{align*}
so we obtain
$$
\begin{aligned}
& \bK \bD^{-1}\bK^\T \\ & = (\bL\bDelta)^{-1} \bPsi_0 \bD \bD^{-1}\bD\bPsi_0^\T \left(\bL\bDelta \right)^{-1}  \\ & = (\bL\bDelta)^{-1} \bPsi_0 \left\{\bPsi_0^\T (\bL\bDelta)^{-1} \bPsi_0\right\}^{-1} \bPsi_0^\T (\bL\bDelta)^{-1}.
\end{aligned}
$$
Furthermore, $\bPsi \bB^{-1} \bPsi^\T = \bPsi(\bPsi^\T \bL\bDelta \bPsi)^{-1} \bPsi^\T$, so using \eqref{eq:matrix3} with $\bB = \bL\bDelta$, we obtain
\begin{equation}
\bPsi \bB^{-1} \bPsi^\T + \bK \bD^{-1}\bK^\T = (\bL\bDelta)^{-1}.
\label{eq:KDK}
\end{equation}
Therefore, equation \eqref{eq:mu} reduces to 
$
\bL^{\T}(\bL\bDelta)^{-1}\bL(\overline{\bW}-\bmu) = \bL^\T \bDelta^{-1}(\overline{\bW}-\bmu) = 0.
$
The matrix $\bL^\T \bDelta^{-1} = (\bDelta + \bSigma_u)^{-1} \bDelta \bDelta^{-1} = (\bDelta + \bSigma_u)^{-1}$ is invertible, so the above equation gives the optimized value $\hat\bmu = \overline{\bW}$. At these optimized values $\hat\bmu$ and $\hat{v}_m$, the log-likelihood is reduced to 
\begin{align*}
& 2 \ell_d  = - \sum_{m=1}^{M} n_m \log \vert \bPsi^\T \bG_m \bPsi \vert  \\ & - n \log \vert \bD\vert \\ & - \sum_{m=1}^{M} n_m \text{tr}\left\{(\bPsi^\T \bL \tilde{\bDelta}_m \bL^\T \bPsi \left(\bPsi^\T \bG_m \bPsi \right)^{-1}\right\}  \\ & - \sum_{m=1}^{M} n_m \text{tr}\left(\bL^\T \bK \bD^{-1} \bK^\T \bL \tilde{\bSigma}_m \right). 
\end{align*}
with $\tilde\bSigma_m = \tilde\bDelta_{wm} + (\overline{\bW}_m - \overline{\bW}) (\overline{\bW}_m - \overline{\bW})^\T$. The maximum likelihood estimator of $\bPsi^\T \bG_m \bPsi$ satisfies $\widehat{\bPsi^\T \bG_m \bPsi} = \bPsi^\T \bL\tilde\bDelta_{wm} \bL^\T \bPsi  $, and therefore, 
\begin{align}
2 \ell_d  = & - \sum_{m=1}^{M} n_m \log \vert \bPsi^\T \bL \tilde{\bDelta}_{wm} \bL^\T  \bPsi  \vert \nonumber \\ & - n \log \vert \bD \vert \nonumber \\ & - \sum_{m=1}^{M} n_m \text{tr}\left(\bL^\T \bK \bD^{-1} \bK^\T \bL \tilde{\bSigma}_m \right). 
\label{eq: D}
\end{align}
Next, we find the maximum likelihood of $\bH$. Since $\bK = \bPsi_0 - \bPsi \bH^\T$, the derivative with respect to $\bH$ equals
$
-\sum_{m=1}^{M} n_m \bD^{-1} \bPsi_0^\T \bL\tilde{\bSigma}_m \bL^\T\bPsi + \sum_{m=1}^{M} n_m \bD^{-1}\bH\bPsi^\T \bL\tilde\bSigma_m \bL^\T \bPsi,
$
which gives the maximum at 
$
\hat{\bH} = (\bPsi_0^\T \bL \tilde\bSigma \bL^\T \bPsi) (\bPsi^\T \bL\tilde\bSigma \bL^\T \bPsi)^{-1}
$
where $\tilde\bSigma \coloneqq \sum_{m=1}^{M} f_m \tilde\bSigma_m$. Hence, $\hat{\bK}^\T = \bPsi_0^\T - \hat{\bH}\bPsi^\T = \bPsi_0^\T (\bI_p -  \bL\tilde\bSigma \bL^\T \bPsi(\bPsi^\T \bL  \tilde\bSigma \bL^\T \bPsi)^{-1}) = \{\bPsi_0^\T (\bL\tilde\bSigma \bL^\T)^{-1}\bPsi_0\}^{-1} \bPsi_0^\T (\bL\tilde\bSigma \bL^\T)^{-1}
$, and the maximum over $\bD$ is 
$$
\begin{aligned}
\widehat{\bD}  &= \hat{\bK}^\T \bL\tilde\bSigma \bL^\T \hat{\bK} \\ & = \{\bPsi_0^\T (\bL\tilde\bSigma_w \bL^\T)^{-1}\bPsi_0\}^{-1} \bPsi_0^\T (\bL\tilde\bSigma_w \bL^\T)^{-1} \bPsi_0  \\ ~ & \times \{\bPsi_0^\T (\bL\tilde\bSigma_w \bL^\T)^{-1}\bPsi_0\}^{-1} \\ &  = \left\{\bPsi_0^\T (\bL\tilde\bSigma_w \bL^\T)^{-1}\bPsi_0\right\}^{-1}.
\end{aligned}
$$
Finally, substituting $\hat{\bD}$ into \eqref{eq: D} and removing terms that do not involve $\bPsi$ and $\bL$, we obtain the joint profile likelihood of $\bPsi$ and $\bDelta$ to be
\begin{equation}
\begin{aligned}
\ell_2(\bPsi, \bDelta)  = & n \log \vert \bPsi_0^\T (\bL\tilde\bSigma_w \bL^\T)^{-1}\bPsi_0 \vert \\ & - \sum_{m=1}^{M} n_m \log  \vert\bPsi^\T \bL \tilde\bDelta_{wm} \bL^\T \bPsi \vert \\ & = -n \log \vert \bL\tilde\bSigma_w \bL^\T \vert  + n \log \vert \bPsi^\T (\bL \tilde \bSigma_w \bL^\T) \bPsi \vert   \\ & - \sum_{m=1}^{M} n_m \log  \vert\bPsi^\T \bL \tilde\bDelta_{wm} \bL^\T \bPsi \vert. 
\end{aligned}
\label{eq:MLE}
\end{equation}

\subsection*{Appendix C: Proof of Theorem 2}
We begin by establishing the equivalence between $\mathcal{S}^*_{\text{cLAD}}$ and $\mathcal{S}^*_{\text{SAVE}}$. First, we establish the equivalence when $\bDelta = \text{E} \left\{\text{Var}(\bX \mid y) \right\}$ (and hence $\bL$) is known, and then we will establish the consistency of the method of moment estimator $\hat\bDelta$. To simplify the notation, we will use $\bDelta_m$ to denote $\bDelta_y^{(m)}$, i.e., the covariance matrix of the inverse Gaussian distribution within the $m$th slice. 

Let $\mathcal{V}_1(\bPsi) = n \log \vert \bPsi^\T (\bL \tilde \bSigma_w \bL^\T) \bPsi \vert - \sum_{m=1}^{M} n_m \log  \vert\bPsi^\T \bL \tilde\bDelta_{wm} \bL^\T \bPsi \vert$, where $\bPsi \in \mathbb{R}^{p \times d}$ is a semi-orthogonal matrix. When $n \rightarrow \infty$,  the function $n^{-1} \mathcal{V}_1(\bPsi)$ converges uniformly to 
$\mathcal{V}_0(\bPsi) = \log \vert \bPsi^\T (\bL\bSigma_w \bL^{\T}) \bPsi \vert - \sum_{m=1}^{M} f_m \log \vert \bPsi^\T (\bL\bDelta_{wm} \bL^{\T}) \bPsi \vert$, 
where $\bDelta_{wm} = \bDelta_m + \bSigma_u$ and $\bSigma_w = \bSigma_x + \bDelta_u$.  Using the result from \citet{cook2009regression} that
$
\vert \bPsi_0^\T \bB \bPsi_0 \vert = \vert \bB \vert \vert \bPsi^\T \bB^{-1} \bPsi \vert 
$  for any symmetric positive semidefinite matrix $\bB$, we obtain
\begin{align*}
    \mathcal{V}_0(\bPsi) & = \log \vert \bPsi_0^\T (\bL\bSigma_w \bL^{\T})^{-1} \bPsi_0 \vert \\ & - \sum_{m=1}^{M} f_m \log \vert \bPsi_0^\T (\bL\bDelta_{wm} \bL^{\T})^{-1} \bPsi_0 \vert + c,
\end{align*}
where $c$ is a quantity that does not depend on $\bPsi$ or $\bPsi_0$. To show the consistency of the estimated central subspace, we show that the true SAVE estimator (i.e defined even when no measurement error is present) is the global maximum for $\mathcal{V}_0(\bPsi)$.

Indeed, as a function of $Q$, the function $\log \vert \bPsi_0^\T Q^{-1} \bPsi_0 \vert$ is convex. As a result, we obtain 
$$
\begin{aligned}
& \sum_{m=1}^{M} f_m \log \vert \bPsi_0^\T (\bL\bDelta_{wm} \bL^{\T})^{-1} \bPsi_0 \vert  \\ & \geq \log \left\vert \bPsi_0^\T \left\{ \bL \left(\sum_{m=1}^{M}f_m \bDelta_{wm}\right)  \bL^\T  \right\}^{-1}  \bPsi_0 \right\vert \\ &= \log \vert \bPsi_0^\T \left\{ \bL(\bDelta + \bSigma_u) \bL^\T \right\}^{-1} \bPsi_0 \vert, 
\end{aligned}
$$
where the second inequality follows from $\sum_{m=1}^{M} f_m = 1$ and $\sum_{m=1}^{M} f_m \bDelta_m = \bDelta$. Next we note that $\bSigma_x = \bDelta + \text{Var}\left\{\text{E}(\bX|y)\right\}$, so $\bSigma_x + \bSigma_u - (\bDelta_x + \bSigma_u) = \bSigma_x - \bDelta $ is still a positive definite matrix. As a result, the matrix difference $ (\bSigma_x + \bSigma_u)^{-1} - (\bDelta + \bSigma_u)^{-1}$ is negative definite, and so is $\bPsi_0^\T \left\{\bL(\bSigma_x +\bSigma_u)\bL^\T \right\}^{-1} \bPsi_0- \bPsi_0^\T \left\{\bL(\bDelta +\bSigma_u)\bL^\T \right\}^{-1}\bPsi_0$. Therefore, we have
\begin{align*}
& \log \vert \bPsi_0^\T \left\{ \bL(\bDelta + \bSigma_u) \bL^\T\right\}^{-1}\bPsi_0 \vert \\ & \geq \log \vert \bPsi_0^\T \left\{\bL(\bSigma_x + \bSigma_u) \bL^\T \right)^{-1} \bPsi_0 \vert \\& = \log \vert \bPsi_0^\T \left(\bL \bSigma_w \bL^\T \right)^{-1}\bPsi_0 \vert.
\end{align*}
As a result, we have $\mathcal{V}_0(\bPsi) \leq c$ and the equality holds only when 
\begin{align}
    & \sum_{m=1}^{M} f_m \log \vert \bPsi_0^\T (\bL \bDelta_{wm} \bL^\T)^{-1} \bPsi_0 \vert \nonumber \\ & = \log \vert \bPsi_0^\T \bL\bSigma_w^{-1} \bL^\T \bPsi_0 \vert.
    \label{eq:SAVE}
\end{align}
We next prove that if $\bm\Phi$ is a semi-orthogonal basis matrix for the SAVE estimator when $\bX$ is observed and $(\bm\Phi, \bm\Phi_0) \in \mathbb{R}^{p \times p}$ is an orthogonal matrix (i.e $\bm\Phi^\T \bm\Phi_0 = 0$), then equation \eqref{eq:SAVE} is satisfied with $\bPsi_0$ replaced by $\bm\Phi_0$. \citet{cook2009likelihood} prove that $\Phi$ satisfies $\bDelta_m^{-1} = \bSigma^{-1} + \bm\Phi \left\{(\bm\Phi^\T \bDelta_m \bm\Phi)^{-1} - (\bm\Phi^\T \bSigma \bm\Phi)^{-1}\right\}\bm\Phi^\T$; as a result, we have $\bm\Phi_0^\T \bDelta_{m}^{-1} = \bm\Phi_0^\T \bSigma^{-1}$. Furthermore, asymptotically, \citet{cook2009likelihood}  prove that the true LAD and SAVE estimator spans the same subspace, and hence $\bm\Phi_0^\T \bDelta_m^{-1} = \bm\Phi_0^\T \bDelta^{-1}$ as well. As a result, we have
$$
\begin{aligned}
& \bm\Phi_0 (\bL\bDelta_{wm}\bL^\T)^{-1} \bm\Phi_0 = \bm\Phi_0 \bL^{-\T} \bDelta_{wm}^{-1} \bL^{-1} \\ & = \bm\Phi_0 \bDelta^{-1} (\bDelta + \bSigma_u)\bDelta_{wm}^{-1} (\bDelta + \bSigma_u) \bDelta^{-1} \bm\Phi_0 \\
& = \bm\Phi_0 (\bI + \bDelta^{-1}\bSigma_u) \bDelta_{wm}^{-1} (\bI + \bSigma_u \bDelta^{-1}) \bm\Phi_0 \\
& = (\bm\Phi_0 + \bm\Phi_0 \bDelta^{-1}\bSigma_u) \bDelta_{wm}^{-1} (\bI + \bSigma_u \bDelta^{-1}) \bm\Phi_0 \\
& =  (\bm\Phi_0 + \bm\Phi_0 \bDelta_{m}^{-1}\bSigma_u) \bDelta_{wm}^{-1} (\bI + \bSigma_u \bDelta^{-1}) \Phi_0 \\ & =  \bm\Phi_0 (\bI + \bDelta_{m}^{-1}\bSigma_u) \bDelta_{wm}^{-1} (\bI + \bSigma_u \bDelta^{-1}) \bm\Phi_0 \\
& = \bm\Phi_0 \bDelta_{m}^{-1} (\bDelta_{m} + \bSigma_u) \bDelta_{wm}^{-1} (\bI + \bSigma_u \bDelta^{-1}) \bm\Phi_0 \\
& = \bm\Phi_0 \bDelta_{m}^{-1} \bDelta_{wm} \bDelta_{wm}^{-1} (\bI + \bSigma_u \bDelta^{-1}) \Phi_0 \\
& = \bm\Phi_0 \bDelta^{-1} (\bI + \bSigma_u \bDelta^{-1}) \bm\Phi_0 ,
\end{aligned}
$$
and 
\begin{equation*}
\begin{aligned}
& \bm\Phi_0 \left(\bL \bSigma_w \bL^\T \right)^{-1} \bm\Phi_0  = \bm\Phi_0 \bL^{-\T} \bSigma_w^{-1} \bL^{-1} \bm\Phi_0  \\
& = \bm\Phi_0 \bDelta^{-1}(\bDelta + \bSigma_u) \bSigma_w^{-1} (\bDelta + \bSigma_u) \bDelta^{-1} \bm\Phi_0 \\
&  = \bm\Phi_0(\bSigma_w^{-1} + \bDelta^{-1} \bSigma_u \bSigma_w^{-1}) (\bI + \bSigma_u \bDelta^{-1}) \bm \Phi_0 \\
& = (\bm\Phi_0^{\T} \bSigma_w^{-1} + \bm\Phi_0 \bDelta^{-1} \bSigma_u\bSigma_w^{-1})(
\bI + \bSigma_u \bDelta^{-1}) \bm\Phi_0  \\ 
& = (\bm\Phi_0^{\T} \bSigma_w^{-1} + \bm\Phi_0 \bSigma^{-1} \bSigma_u\bSigma_w^{-1})(\bI + \bSigma_u \bDelta^{-1}) \bm\Phi_0 \\
& = \bm\Phi_0 \bSigma^{-1}(\bSigma+\bSigma_u)\bSigma_w^{-1}(\bI + \bSigma_u \bDelta^{-1}) \bm\Phi_0 \\
& = \bm\Phi_0 \bSigma^{-1}\bSigma_w\bSigma_w^{-1}(\bI + \bSigma_u \bDelta^{-1}) \bm\Phi_0 \\
& = \bm\Phi_0 \bDelta^{-1}(\bI + \bSigma_u \bDelta^{-1}) \bm\Phi_0 ,
\end{aligned}
\end{equation*}
which verifies \eqref{eq:SAVE} for the SAVE estimator. 

Finally, since we estimate $\bL$ by $\hat{\bL}$ from the naive LAD estimator, we will prove that $\hat{\bL}$ is consistent for $\bL$. It suffices to prove that $\hat\bDelta_\text{n}$ is a consistent estimator of $\text{E}\left\{\text{Var}(\bW \mid y\right\}$. By the relationship between LAD and SAVE, we have $\hat\bPsi_\text{n}$ converges to a naive SAVE population $\Phi_{\text{n}}$ that satisfies
\begin{equation*}
\begin{aligned}
(\bSigma_u + \bDelta_{m})^{-1}  &= (\bSigma_x+ \bSigma_u)^{-1} \\ &  + \bPhi_{\text{n}} \left[\left\{\bPhi_\text{n}^\T (\bDelta_{m} + \bSigma_u) \bPhi_\text{n}\right\}^{-1} \right. \\  & \left. - \left\{\bPhi_\text{n}^\T (\bSigma_x + \bSigma_u)  \bPhi_\text{n}\right\}^{-1}\right]\bPhi_\text{n}^\T,
    \end{aligned}
\end{equation*}
and
\begin{align*}
(\bSigma_u + \bDelta_{m})^{-1}  & = (\bSigma_u + \bDelta)^{-1}  \\ & +  \bPhi_{\text{n}}\left\{\bPhi_{\text{n}}^\T \left(\bDelta_{m} + \bSigma_u\right)\bPhi_{\text{n}}\right\}^{-1} \bPhi_{\text{n}}^\T  \\ &-  \bPhi_\text{n} \left\{\bPhi_\text{n}^\T (\bDelta +  \bSigma_u) \bPhi_\text{n}\right\}^{-1} \bPhi_\text{n}^\T.
\end{align*}
Therefore, as $n \to \infty$, we have
\begin{align*}
\hat\bDelta_\text{n}^{-1} = 
& \hat\bPsi_\text{n}(\hat\bPsi_\text{n}^\T \tilde\bDelta\hat\bPsi_\text{n})^{-1} \hat\bPsi_\text{n}^\T  + \tilde\bSigma_w^{-1} ] \\ &  - \hat\bPsi_\text{n}(\hat\bPsi_\text{n}^\T \tilde\bSigma_w\hat\bPsi_\text{n})^{-1} \hat\bPsi_\text{n}^\T 
\end{align*} 
converges to 
$$
\begin{aligned}
& \bPhi_{\text{n}}\left\{\bPhi_{\text{n}}^\T \left(\bDelta_{m} + \bSigma_u\right)\bPhi_{\text{n}}\right\}^{-1} \bPhi_{\text{n}}^\T  + (\bSigma_x + \bSigma_u)^{-1} \\ & - \bPhi_{\text{n}}\left\{\bPhi_{\text{n}}^\T (\bSigma_x + \bSigma_u)\bPhi_{\text{n}}\right\}^{-1} \bPhi_{\text{n}}^\T \\
& = (\bSigma_u + \bDelta_{m})^{-1} + \bPhi_{\text{n}}\left\{\Phi_{\text{n}}^\T \left(\bDelta_{m} + \bSigma_u\right)\bPhi_{\text{n}}\right\}^{-1} \bPhi_{\text{n}}^\T \\ & -  \bPhi_\text{n} \left\{\bPhi_\text{n}^\T (\bDelta + \bSigma_u) \bPhi_\text{n}\right\}^{-1} \bPhi_\text{n}^\T 
\\ & = (\bSigma_u + \bDelta)^{-1}, 
\end{aligned}
$$
so $\hat\bDelta_n$ converges to $\bSigma_u + \bDelta = \text{E}\left\{\text{Var}(\bW \mid y) \right\}$. The proof is now complete. 

To prove the equivalence between $\mathcal{S}^*_{\text{IL--LAD}}$ and $\mathcal{S}^*_{\text{SAVE}}$, by a similar argument, it suffices to establish
\begin{equation}
\sum_{m=1}^{M} f_m \log \vert \bm\Phi_0^\T \left(\bm\Delta_m^* \right)^{-1} \bm\Phi_0 \vert = \log \vert \bm\Phi_0^\T \left(\bm\Sigma^* \right)^{-1} \bm\Phi_0 \vert, 
\label{eq:SAVE2}
\end{equation}
where $\bm\Phi_0$ is the orthogonal complement of the semi-orthogonal basis $\bm\Phi$ corresponding to the SAVE estimator, and with $\bm\Sigma^* = \bm\Sigma_x\bm\Sigma_w^{-1}\bm\Sigma_x$, and $\bm\Delta_m^* = \bm\Sigma_x \bm\Sigma_{w}^{-1}\left( \bm\Delta_m+ \bm\Sigma_u\right)\bm\Sigma_{w}^{-1}\bm\Sigma_x$. Since $\bm\Phi_0^\T \bm\Sigma_x^{-1} = \bm\Phi_0^\T \bm\Delta_m^{-1}$,  we have 
$$
\begin{aligned}
&\bm\Phi_0^\T \left(\bm\Delta_m^* \right)^{-1} \bm\Phi_0 \\ &= \bm\Phi_0^\top \bm\Sigma_x^{-1} \bm\Sigma_{w} \left( \bm\Delta_m+ \bm\Sigma_u\right)^{-1}\bm\Sigma_{w}\bm\Sigma_x^{-1}\bm\Phi_0 \\
& = \bm\Phi_0^\top \bm\Sigma_x^{-1} (\bm\Sigma_{x} + \bm\Sigma_u) \left( \bm\Delta_m+ \bm\Sigma_u\right)^{-1}\bm\Sigma_{w}\bm\Sigma_x^{-1}\bm\Phi_0 \\
& = (\bm\Phi_0^\top +  \bm\Phi_0^\top\bm\Sigma_x^{-1} \bm\Sigma_u) \left( \bm\Delta_m+ \bm\Sigma_u\right)^{-1}\bm\Sigma_{w}\bm\Sigma_x^{-1}\bm\Phi_0 \\
& = (\bm\Phi_0^\top +  \bm\Phi_0^\top\bm\Delta_m^{-1} \bm\Sigma_u) \left( \bm\Delta_m+ \bm\Sigma_u\right)^{-1}\bm\Sigma_{w}\bm\Sigma_x^{-1}\bm\Phi_0 \\
& = \bm\Phi_0^\top \bm \Delta_m^{-1}(\bm\Delta_m + \bm\Sigma_u) \left( \bm\Delta_m+ \bm\Sigma_u\right)^{-1}\bm\Sigma_{w}\bm\Sigma_x^{-1}\bm\Phi_0 \\
& = \bm\Phi_0^\top \bm \Sigma_x^{-1} \bm\Sigma_{w}\bm\Sigma_x^{-1}\bm\Phi_0 \\
& = \bm\Phi_0^\top  \bm \left( \bm\Sigma^* \right)^{-1} \bm\Phi_0,
\end{aligned}
$$
so \eqref{eq:SAVE2} follows from $\sum_{m=1}^{M} f_m = 1$. 

\bibliographystyle{apalike}
\bibliography{arxiv}
\end{document}